\documentstyle[preprint,eqsecnum,aps,epsfig]{revtex}

\newtheorem{lem}{Lemma}[section]
\newtheorem{preremark}{Remark}[section]

\newtheorem{defi}{Definition}[section]
\newtheorem{theorem}{Theorem}[section]
\newtheorem{prop}{Proposition}[section]

\newcommand{\bequ}[1]{\begin{equation}\label{#1}}
\newcommand{\eequ}{\end{equation}}
\newcommand{\barr}[1]{\begin{eqnarray}\label{#1}}
\newcommand{\earr}{\end{eqnarray}}
\newcommand{\barrz}{\begin{eqnarray*}}
\newcommand{\earrz}{\end{eqnarray*}}
\newcommand{\Rf}[1]{(\ref{#1})}
\newcommand{\SSS}{{S^{{\rm 1}}}}
\newcommand{\RR}{{\bf {R}}}
\newcommand{\ZZ}{{\bf {Z}}}
\newcommand{\QQ}{{\bf {Q}}}
\newcommand{\RRR}{ {\bf R}^2}
\newcommand{\CC}{{\bf {C}}}
\newcommand{\NN}{{\bf {N}}}
\newcommand{\OPH}{orientation preserving homeomorphism}

\newcommand{\mod}{\mathop{\rm mod}\nolimits}
\newcommand{\id}{\mathop{\rm Id}\nolimits}
\newcommand{\der}{{\rm d}}
\newcommand{\differ}[1]{{{#1}'}}
\newcommand{\fal}{r_{\alpha}}
\newcommand{\Fal}{R_{\alpha}}

\newcommand{\Gtil}{{\widetilde{G}}}

\newcommand{\Heps}{{H_\varepsilon}}
\newcommand{\heps}{{h_\varepsilon}}
\newcommand{\htil}{{\eta}}

\newcommand{\calN}[2]{{{N}_{#1}(#2)}}
\newcommand{\addition}{{b}}

\newcommand{\psiminus}{{\Psi^{-}}}
\newcommand{\psiplus}{{\Psi^{+}}}
\newcommand{\psipm}{{\Psi^{\pm}}}

\newcommand{\proofend}{\vspace{3mm}}

\begin{document}

\draft

\title{Theory of Circle Maps
	and the Problem of One-Dimensional Optical 
	Resonator with a Periodically Moving Wall}
\author{Rafael de la Llave\cite{email_rl}}
\address{Department of Mathematics \\
	University of Texas at Austin \\
Austin, TX 78712, USA}
\author{Nikola P. Petrov\cite{email_np}}
\address{Department of Physics \\
	University of Texas at Austin \\
Austin, TX 78712, USA}
\date{October 7, 1998}
\maketitle
\begin{abstract}
We consider the electromagnetic field in a cavity 
with a periodically oscillating 
perfectly reflecting boundary 
and show that the mathematical 
theory of circle maps leads 
to several physical predictions. 
Notably, well-known results 
in the theory of circle maps 
(which we review briefly) 
imply that there are intervals 
of parameters where the waves in the 
cavity get concentrated in wave packets 
whose energy grows exponentially. 
Even if these intervals are dense 
for typical motions of the reflecting boundary,  
in the complement there is a 
positive measure set of parameters 
where the energy remains bounded.
\end{abstract}

\narrowtext


\section{INTRODUCTION}

In this paper, we consider the behavior 
of the electromagnetic 
field in a resonator one of whose walls is at rest 
and the other moving periodically. 
The main point we want to make 
is that several results 
in the mathematical literature 
of circle maps immediately yield 
physically important conclusions. 

The problem at hand is mathematically equivalent to 
the study of the motion of vibrating strings 
with a periodically moving boundary 
\cite{Balazs1961,Dittrichetal1997}, 
or the classical electromagnetic field 
in a periodically pulsating cavity 
\cite{Cooper1993b,ColeSchieve1995}. It is connected 
with the vacuum quantum effects in such region 
\cite{Moore1970,Law1994a}. 
The problem is also of practical importance, 
e.g., for the formation of short laser pulses 
\cite{HennebergerSchulte1966}.

The goal of this paper is to show that the problem 
of a classical wave with a periodically moving boundary 
can be easily reformulated 
in terms of the study of long term 
behavior of circle maps and, 
therefore, that many well known results 
in this theory lead to 
physical predictions. 
In particular, we give proofs of several 
results obtained numerically 
by Cole and Schieve \cite{ColeSchieve1995} and others. 
Extensions of this approach, that will be discussed
elsewhere, allow to reach conclusions for
some quasi-periodic motions of small 
amplitude or possibly for non-homogeneous media.

In the case of more than one spatial dimension, 
the analogous problem 
\cite{CooperStrauss1976} 
is much more complicated, 
so the predictions are not as clear 
as in the one-dimensional case and 
we will not discuss them further. 
 
We emphasize that the mathematical 
theory presented is completely
rigorous and, hence, the physical predictions made 
are general for the assumptions stated.

There are other intriguing relations for which 
we have no conceptual explanation.
We observe that a calculation 
of Fulling and Davies \cite{FullingDavies1976} 
leads to the conclusion that 
the energy density radiated by a moving mirror 
is equal to the Schwarzian derivative of
the motion of the mirror 
(for details see Sec.~\ref{sec:schwarz}). 
This Schwarzian derivative is a differential 
operator frequently used in the theory of 
one dimensional dynamical systems 
and particularly in the theory of circle maps.

The plan of the exposition is the following. 
In Sec.~\ref{phys_set} we show how 
the physical problem can be formulated 
in terms of circle maps.
Sec.~\ref{sec:maps-cir} contains 
a brief exposition of the necessary 
facts from the theory of circle maps, 
in Sec.~\ref{sec:appl-th} 
these facts are applied to the problem 
at hand and illustrated numerically, 
and in the conclusion we discuss 
the advantages of our approach.

\section{Physical setting}
\label{phys_set}

\subsection{Description of the system}
\label{sec:descr_system}

We consider a one-dimensional 
optical resonator consisting of 
two parallel perfectly reflecting mirrors. 
For simplicity of notation, 
we will consider only the situation 
in which one of them is at rest 
at the origin of the $x$ axis while 
the other one is moving periodically 
with period $T$. The case where 
the two mirrors are moving periodically 
with a common period can be treated 
in a similar manner.
We assume that the resonator is empty, 
so that the speed of the electromagnetic waves 
in it is equal to the speed 
of light, $c$. The speed of the moving 
mirror cannot exceed~$c$.

We note that the experimental 
situation does not necessarily 
require that there is a physically moving mirror. 
One experimental possibility -- among others -- 
would be to have a material 
that is a good conductor or not depending on 
whether a magnetic field of sufficient intensity
is applied to it, and then 
have a magnetic field applied to it 
in a changing region. 
This induces reflecting boundaries 
that are moving with time. 
Note that the boundaries of this region 
could move even faster than~$c$, 
hence the study of  
mirrors moving at a speed comparable to $c$ 
is not unphysical 
(even if, presumably, one would also have to 
discuss corrections to the boundary conditions 
depending on the details 
of the experimental realizations).

We shall use dimensionless time $t$ 
and length $\ell$ connected with 
the physical (i.e., dimensional) time $t_{\rm phys}$ 
and length $\ell_{\rm phys}$ by 
$t := t_{\rm phys}/T$, $\ell := \ell_{\rm phys}/(cT)$.

Let the coordinate of the moving 
mirror be $x=a(t)$, where $a$ 
is a $C^k$ function 
($k=1,\,\ldots,\,\infty,\,\omega$) 
satisfying the conditions
\begin{equation}
\label{properties_a}
a(t)>0  \,, \qquad |\differ{a}(t)| < 1 
	\,, \qquad a(t+1)=a(t) \,.
\end{equation}
The meaning of the first condition is that the cavity 
does not collapse, the second one means that 
the speed of the moving mirror cannot 
exceed the speed of light, and the 
third one is that the mirror's motion 
is periodic of period~1. 
An example which we will use for 
numerical illustrations is
\begin{equation}
\label{equ:a1}
a(t) = \frac{\alpha}{2} + {\beta} \sin{2\pi t} \qquad
\left( |\beta| < \frac{1}{2\pi}\,, \quad 
0 < |\beta| < \frac{\alpha}{2} \right) \,.
\end{equation}

Since there are no charges and no currents, we impose the 
gauge conditions $A_0=0$, $\nabla\cdot{\bf A}=0$
on the 4-potential $A_\mu=(A_0, {\bf A})$ 
and obtain that ${\bf A}$ satisfies 
the homogeneous wave equation. 
We consider plane waves traveling 
in $x$-direction, so that without loss 
of generality, we assume that
${\bf A}(t,x) = A(t,x)\,{\bf e}_y$, 
and obtain that $A(t,x)$ must satisfy 
the homogeneous $(1+1)$-dimensional wave equation, 
\begin{equation}
\label{eq:eq}
A_{tt}(t,x) - A_{xx}(t,x) = 0 \,, 
\end{equation}
in the domain 
$\Sigma := \{(t,x)\in\RRR \,|\, t_0 < t, 
		\,\, 0<x<a(t) \}$.
It will also need to satisfy some boundary 
conditions that will be specified 
in Sec.~\ref{sec:met_charact}, 
and appropriate initial conditions, 
\begin{equation}
\label{eq:ic}
A(t_0,x) = \psi_1(x) \,, \qquad A_t(t_0,x) 
	= \psi_2(x) \,.
\end{equation}

Before discussing the boundary conditions 
and the method of solving 
the boundary-value problem in the domain $\Sigma$, 
let us discuss the way of solving \Rf{eq:eq} 
in the absence of spatial boundaries, i.e., in the domain 
$\{t_0 < t,\,x\in\RR\}$. 
It is well-known that in this case, 
the solution of the problem 
\Rf{eq:eq}, \Rf{eq:ic} 
at some particular space-time point $(t,x)$ can be written as 
\begin{equation}
\label{eq:char_space}
A(t,x) = \psiminus (x_0^-) + \psiplus (x_0^+)	\,,
\end{equation}
where $x_0^\pm := x \pm (t - t_0)$, and 
$\psiminus$ and $\psiplus$ are functions of one variable 
that are selected to match the initial conditions \Rf{eq:ic}. 
The explicit expressions for $\psipm$ 
follow from the d'Alembert's 
formula (see, e.g.,~\cite{John1982})
\begin{equation}
\label{eq:fpm}
\psipm (s) = \frac12 \left[ \psi_1(s) 
		\pm \int_\zeta^s \psi_2(s') 
		\, \der s' \right]	\,,
\end{equation}
where $\zeta$ is an arbitrary constant 
(the same for $\psiplus$ and $\psiminus$).

The representation \Rf{eq:char_space} 
has a simple geometrical meaning:  
the value of $A(t,x)$ is a superposition of 
two functions, $\psiminus (x_0^-)$ 
and $\psiplus (x_0^+)$, 
the former being constant along 
the lines $\{x-t={\rm const}\}$, 
and the latter being constant 
along $\{x+t={\rm const}\}$. 
The disturbances at a space-time 
point $(T,X)$ propagate 
in the space-time diagram along 
the lines 
$\{x-t=X-T\}$ and $\{x+t=X+T\}$ 
emanating from this point 
(in more physical terms, this 
corresponds to two rays moving to the right 
and to the left at unit speed); 
these lines are called {\em characteristics}, 
and the method of solving \Rf{eq:eq}, \Rf{eq:ic} 
by using the representation \Rf{eq:char_space} 
is called the {\em method of characteristics} 
(see, e.g., \cite{John1982,Garabedian1964}).


\subsection{Method of characteristics,
	boundary condition at the moving mirror, 
	and Doppler shift at reflection}
\label{sec:met_charact}

To obtain the boundary conditions at the stationary mirror, 
we note that the electric field, 
i.e., the temporal derivative 
of the vector potential, must vanish 
at this mirror, which yields the following 
``perfect reflection'' boundary condition:
\begin{equation}
\label{eq:bc1}
A_t(t,0)=0	\,.
\end{equation}

The boundary condition at the moving mirror 
can be easily obtained by performing 
a Lorenz transformation from the laboratory frame 
to the inertial frame comoving with the moving mirror 
at some particular moment~$t$. 
The temporal and spatial coordinates 
in the laboratory frame, $t$ and $x$, 
are related to the ones in the 
instantaneously comoving inertial frame, 
$t'$ and $x'$, by
\begin{eqnarray}
\label{eq:Lorenz}
t - t_0 &=& t' \cosh \zeta + x' \sinh \zeta 
			\nonumber \\ [-2mm]
			\\ [-2mm]
x - a(t_0) &=& t' \sinh \zeta 
	+ t' \cosh \zeta \,, \nonumber
\end{eqnarray}
where $\tanh\zeta=a'(t)$. 
In the comoving frame, the boundary condition is 
$A_{t'}(0,0)=0$, which, together 
with~\Rf{eq:Lorenz}, yields 
\begin{equation}
\label{eq:bc2}
A_t(t,a(t)) + a'(t) \, A_x(t,a(t)) = 0	\,.
\end{equation}

The method of characteristics 
developed in \Rf{eq:char_space} and \Rf{eq:fpm} 
for situations with no boundaries 
can be adapted to provide rather explicit 
solutions for systems in 
spatially bounded space-time domains 
satisfying \Rf{eq:bc2} at the boundaries 
(see, e.g., \cite{Weinberger1965}).

\begin{figure}
\centerline{\epsfig{file=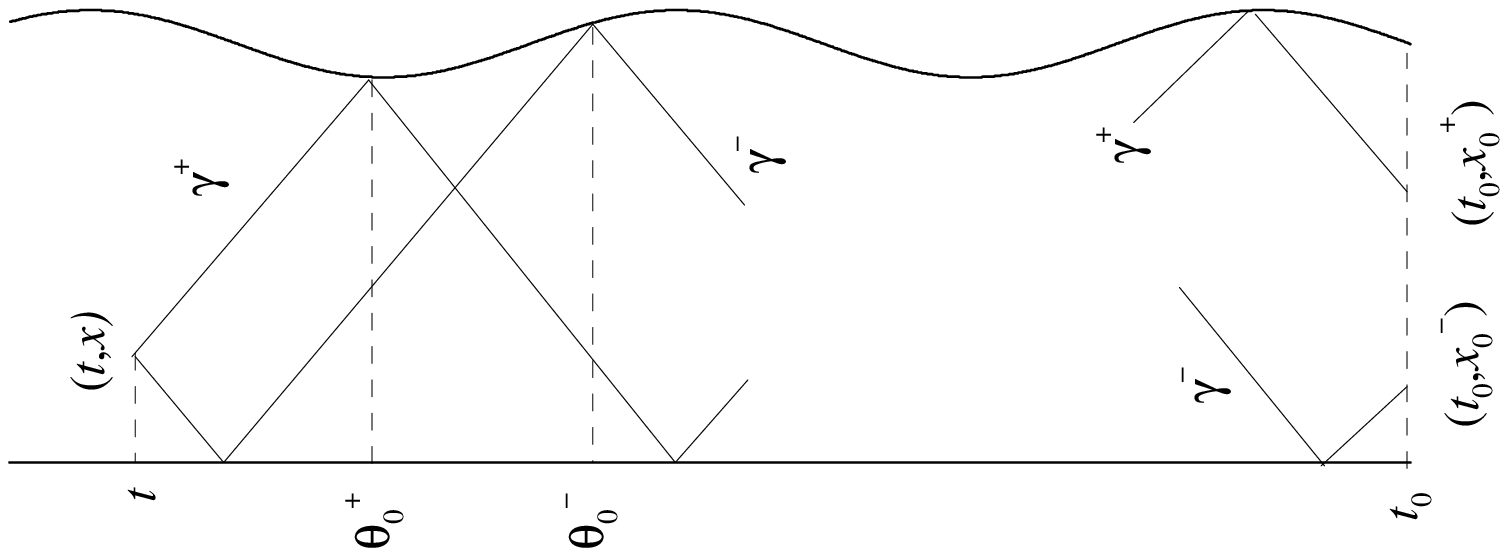, height= 2 in, angle=-90}} 
\bigskip
\caption{Finding $A(t,x)$ by the method of characteristics.}
\label{fig:charact_dop}
\end{figure}

The prescription is the following. 
The solution of the boundary value problem 
\Rf{eq:eq}, \Rf{eq:ic}, \Rf{eq:bc1}, \Rf{eq:bc2} 
in the domain $\Sigma$ 
is a superposition of two functions that are constant 
on the straight pieces of the characteristics 
and change their sign at each reflection. 
To find $A(t,x)$, one has to 
consider the two characteristics, 
$\gamma^-$ and $\gamma^+$, 
passing through $(t,x)$, 
and propagate them backwards in time 
(according to the rule that, upon reaching a mirror, 
they change direction of propagation) 
until they reach the line $\{ {\rm time} = t_0 \}$ 
at the points $(t_0,x_0^-)$ and $(t_0,x_0^+)$, resp.\ 
-- see Fig.~\ref{fig:charact_dop}. 
Then $A(t,x)$ is given by 
\begin{equation}
\label{eq:sol_sigma}
A(t,x)	= (-1)^{N_{-}} \psiminus(x_0^{-}) 
	+ (-1)^{N_{+}} \psiplus(x_0^{+})	\,,
\end{equation}
where $N_\mp$ are the number of reflections 
of $\gamma^\mp$ on the way back from $(t,x)$ 
to $(t_0,x_0^\mp)$. 
In Sec.~\ref{sec:sol_problem} we will 
give explicit formulae for $x_0^\mp$ and $A(t,x)$ 
in terms in circle maps.

Indeed, because the solution \Rf{eq:sol_sigma} 
is the sum of two functions constant along the 
straight pieces of the characteristics, 
the wave equation is satisfied in the interior. 
Also, the initial conditions are easily verified 
because for $t-t_0$ small, 
$x_0^-$ and $x_0^+$ are close to $x$ 
[see \Rf{eq:x0pm}].

\begin{figure}
\centerline{\epsfig{file=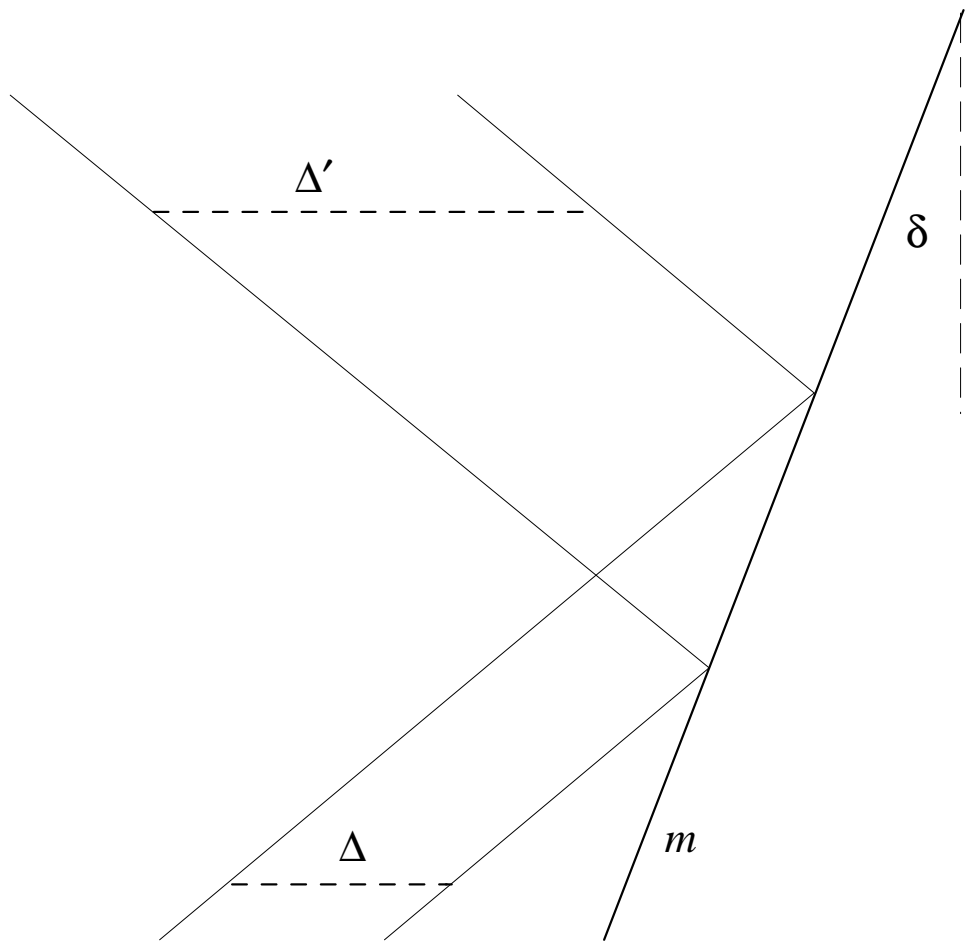, height= 5 in, angle=-90}} 
\bigskip
\caption{Reflection by the moving mirror.}
\label{fig:reflection}
\end{figure}

To check that this prescription 
also satisfies the boundary conditions, 
we need another argument. 
Consider the space-time diagram of the reflection 
of the field between two infinitesimally close 
characteristics reflected by the moving mirror 
at time~$\theta$, shown in Fig.~\ref{fig:reflection}. 
The world line of the mirror is denoted by~$m$, 
the angle $\delta$ between it 
and the time direction is connected 
with the mirror's velocity at reflection by 
$\tan\delta=\differ{a}(\theta)$. 
The Doppler factor at reflection, $D(\theta)$, 
is defined as the ratio of the spatial distances 
$\Delta$ and $\Delta'$ between the characteristics 
before and after reflection: 
\begin{equation}
\label{Doppler_factor}
D(\theta):=\frac{\Delta}{\Delta'}
	= \tan\left(\frac{\pi}{4} - \delta\right) 
	= \frac{1-\tan\delta}{1+\tan\delta}
	= \frac{1-\differ{a}(\theta)}
		{1+\differ{a}(\theta)}	\,.
\end{equation}
Thus, the absolute values of the temporal 
and spatial derivatives of the field 
increase by a factor of $D(\theta)$ 
after reflection. This implies that 
if in the space-time domain between 
the two characteristics, the values of 
the corresponding derivatives 
of the field before reflection are denoted 
by ${\sf A}_t$ and ${\sf A}_x$, 
then after reflection they will become 
$-D(\theta){\sf A}_t$ and 
$D(\theta){\sf A}_x$, resp. 
Hence, in the space-time domain of the overlap 
the derivatives of the field will be 
\begin{eqnarray}
\label{eq:der's}
A_t(\theta,a(\theta))
	&= {\sf A}_t - D(\theta){\sf A}_t
			\nonumber	\\ [-2mm]
					\\ [-2mm]
A_x(\theta,a(\theta))
	&= {\sf A}_x + D(\theta){\sf A}_x	\,. 
			\nonumber
\end{eqnarray} 
Now, we will show that 
the modified method of characteristics 
is consistent with the boundary condition~\Rf{eq:bc2}. 
We note that ${\sf A}_t=-{\sf A}_x$, 
which simply means that before reflection 
the rays are moving to the right at unit speed. 
If we multiply the second equation of \Rf{eq:der's} 
by $a'(\theta)=\frac{1-D(\theta)}{1+D(\theta)}$ 
[which follows from \Rf{Doppler_factor}] and 
add it to the first, we obtain exactly 
the boundary condition \Rf{eq:bc2}.

%
%
%
%

The same prescription gives a solution 
of the Dirichlet problem $A(t,0)=A(t,a(t))=0$. 
Similar methods can be developed 
for other boundary conditions.

We note that the method of characteristics 
also yields information in the important 
case when the medium is inhomogeneous  and 
perhaps time dependent. This is a physically
natural problem since in many applications
we have cavities 
filled with optically active media 
whose characteristics are changed
by external perturbations. 
In this case, the method of characteristics 
does not yield an exact solution as above
but rather, it is the main ingredient 
of an iterative procedure \cite{Garabedian1964}. 
Physically, what happens is that in 
inhomogeneous media, the waves change shape 
while propagating 
in contrast with  the 
propagation without change in shape 
in homogeneous media~(\ref{eq:char_space}).
We plan to come to this problem in 
a near future.


\subsection{Using circle maps to solve 
	the boundary-value problem 
	\Rf{eq:eq}, \Rf{eq:ic}, \Rf{eq:bc1}, \Rf{eq:bc2} 
	in the domain $\Sigma$}
\label{sec:sol_problem}

We now reformulate the method of characteristics 
into a problem of circle maps.

We consider a particular characteristic 
and denote by $\{\tau_n\}$ the times 
at which it reaches the stationary mirror 
and $\{\theta_n\}$ the times at which 
it reaches the oscillating one; let 
$\ldots < \tau_n < \theta_n < \tau_{n+1} 
	< \theta_{n+1} < \ldots$. 
Note that, with this notation, 
\begin{eqnarray}
\label{eq:theta_tau}
\tau_{n} 	&=& \theta_n - a(\theta_n) 
		= (\id - a) (\theta_n)
			\nonumber	\\ [-1.5mm]
				\\ [-1.5mm]
\tau_{n+1}	&=& \theta_n + a(\theta_n) 
		= (\id + a) (\theta_n)	
			\nonumber\,. 
\end{eqnarray} 
Therefore
\begin{eqnarray}
\label{eq:Fexp}
\tau_{n+1} &=& (\id + a) \circ (\id - a)^{-1} (\tau_n)
		=: F(\tau_n)	\nonumber	\\ [-1.5mm]
						\\ [-1.5mm]
\theta_{n+1} &=& (\id - a)^{-1} \circ (\id + a) (\theta_n) 
		=: G(\theta_n)	\,.	\nonumber
\end{eqnarray} 
We refer to $F$ and $G$ as the time advance maps. 
They allow to compute the time of reflection 
on one side in terms of the time 
of the previous reflection on the same side. 
The conditions \Rf{properties_a} 
on the range of $a$ and $\differ{a}$ 
guarantee that $(\id-a)$ is invertible 
and that $F$ and $G$ are $C^k$ 
(by the implicit function theorem). 

When the function $a$ is 1-periodic, 
$F$ and $G$ satisfy
\bequ{eq:Fper}
F(t+1) = F(t) + 1	\,,	\qquad
G(t+1) = G(t) + 1	\,.
\eequ
These relations mean that $F(t)$ and $G(t)$ 
depend only on the fractional part of $t$.
In physical terms, we characterize 
a reflection of a ray
by the {\em phase} of the oscillating mirror 
when the impact takes place, 
i.e., by the time of reflection modulo~1; 
if we know the phase 
at one reflection, we can compute 
the phase at the next impact.
Mathematically, this means that $F$ and $G$ 
can be regarded as lifts of maps from 
$\SSS\equiv\RR/\ZZ$ to $\SSS$ 
(see Sec.~\ref{sec:maps-cir}).

We want to argue that the study 
of the dynamics of the circle maps 
(\ref{eq:Fexp}) leads to important conclusions 
for the physical problem, which we will take 
up after we collect some  information about the 
mathematical theory of circle maps. 
In particular, many results 
in the mathematical literature 
are directly relevant for physical applications.
This is natural because the long term behavior 
of the solution can be obtained by repeated 
application of the time advance maps 
[see \Rf{eq:vect_pot}].

We call attention to the fact that
\begin{equation}
\label{conjFG}
G 	= (\id + a)^{-1} \circ F \circ (\id + a)
	= (\id - a)^{-1} \circ F \circ (\id - a)  \,, 
\end{equation}
so that 
$$
G^n 	= (\id + a)^{-1} \circ F^n \circ (\id + a)
	= (\id - a)^{-1} \circ F^n \circ (\id - a)  \,.
$$
In dynamical systems theory this is usually 
described as saying that 
the maps $F$ and $G$ are ``conjugate'' 
(see Sec.~\ref{sec:PoincareDenjoy}). 
In our situation, this comes from 
the fact that $F$ and $G$ are 
physically equivalent descriptions 
of the relative phase of 
different successive reflections: 
$F$ advances the $\tau$ variables 
while $G$ advances the $\theta$'s, 
and the $\theta$'s are related 
to the $\tau$'s by \Rf{eq:theta_tau}.

Now, we use circle maps to derive 
an explicit formula for the solution 
of the boundary-value problem 
\Rf{eq:eq}, \Rf{eq:ic}, \Rf{eq:bc1}, 
\Rf{eq:bc2} in the domain $\Sigma$. 
Let us trace back in time the characteristics 
$\gamma^-$ and $\gamma^+$ coming ``from the past'' 
to the space-time point $(t,x)$ 
-- see Fig.~\ref{fig:charact_dop}. Let 
$
\theta^\pm_0 := (a\mp\id)^{-1} (t-x)
$
be the last moments the characteristics $\gamma^\pm$ 
are reflected by the moving mirror, and let 
$
\theta^\pm_{-k} := G^{-k} (\theta_0^\pm) 
$.
After $N_+$, resp.\ $N_-$, 
reflections on the way backwards in time 
(out of which $n_+$, resp.\ $n_-$, 
are from the moving mirror), 
the characteristic $\gamma^+$, 
resp.\ $\gamma^-$, crosses the line $\{{\rm time}=t_0\}$. 
The spatial coordinate of the intersection of 
$\gamma^\pm$ and $\{{\rm time}=t_0\}$ can be easily seen to be
\begin{equation}
\label{eq:x0pm}
x_0^\pm = h (\theta_{-n_\pm}^\pm,\,t_0) := 
\bigl| (\id - a) (\theta_{-n_\pm}^\pm) - t_0 \bigr|	\,.
\end{equation}
Thus, the formula for the vector potential is
\begin{eqnarray}
\label{eq:vect_pot}
A(t,x)	&=& (-1)^{N_-} \psiminus\circ 
		h\left(G^{-n_{-}}\circ(a+\id)^{-1}
		(t-x),\,t_0\right)	\nonumber \\[1mm]
	& & + (-1)^{N_+} \psiplus\circ 
		h\left(G^{-n_{+}}\circ(a-\id)^{-1}
			(t-x),\, t_0\right)	\,.
\end{eqnarray}


\subsection{Energy of the field}
\label{sec:Dop_shift}

The method of characteristics 
gives a very illuminating picture 
of the mechanism of the change 
of the field energy, 
\begin{equation}
\label{eq:energy}
E(t) = \int_0^{a(t)} {\cal T}^{00} (t,x) 
	= \frac{1}{8\pi} \int_0^{a(t)} 
	\left[ A_t(t,x)^2 + A_x(t,x)^2 \right] \, \der x  \,,
\end{equation}
due to the distortion of the wave 
at reflection from the moving mirror.
Indeed, consider the change of the energy 
of a very narrow wave packet at reflection 
from the moving mirror at time $\theta$. 
Since at reflection the temporal and the spatial 
distances decrease by a factor of $D(\theta)$, 
$|A_t|$ and $|A_x|$ will increase 
by a factor of $D(\theta)$. 
Therefore, the integrand of the energy integral 
will increase $D(\theta)^2$ times, 
while the support of the integrand 
(i.e., the spatial width of the 
wave packet at time~$t$) 
will shrink by a factor of $D(\theta)$. 
Hence, the energy of the wave packet 
after reflection will be $D(\theta)$ times 
greater than its energy before reflection.

In the general case, one can use \Rf{eq:vect_pot} and 
obtain the energy of the system at time~$t$. 
For the sake of simplicity, 
we will give the formula only under the assumption that 
at time~$t$ all the rays are going to the right, 
i.e., assuming that the vector potential is of the form 
$A(t,x) = (-1)^{N_-} \psiminus(x_0^-)$. 
Let us introduce the ``local Doppler factor'' 
\begin{equation}
\label{eq:locDopfac}
D(t_0,\,x_0^-;\,t) := \left| \frac{\partial}{\partial t} 
		h(\theta_{n_{-}}^{-},\,t_0) \right|	
	= \frac{1-a'(\theta^-_{-n_-})}{1+a'(\theta^-_0)} \,
		(G^{-n_-})'(\theta_0^{-})	\,.
\end{equation}
It has the physical meaning of the ratio of the frequencies 
of the incident wave and the wave at time~$t$ 
[cf.~\Rf{Doppler_factor}]. 
Note that $D(t_0,\,x_0^-;\,t)$ 
is equal to the derivative of $G^{-n_-}$ 
multiplied by a factor which is bounded and 
bounded away from~0 independently of $n_-$ 
[due to the fact that $|a'(t)|<1$]. 
From \Rf{eq:vect_pot} and \Rf{eq:x0pm} we obtain 
that the square of $D(t_0,\,x_0^-;\,t)$ is the 
ratio of the energy density ${\cal T}^{00} (t,x)$ 
and the initial energy density, ${\cal T}^{00} (t_0,x_0^-)$: 
$$
{\cal T}^{00} (t,x) 
	= 2\,\left|(\psiminus)'(x_0^-)\right|^2 \,
		D(t_0,\,x_0^-;\,t)^2 
	= {\cal T}^{00} (t_0,x_0^-) \, 
		D(t_0,\,x_0^-;\,t)^2	\,.
$$
On the other hand, $D(t_0,\,x_0^-;\,t)$ 
is connected with the Jacobian 
of the change of coordinates $x_0^- \mapsto x$ by 
$$
\left| \frac{\partial x}{\partial x_0^-} \right| 
	= \left| \frac{\partial x_0^-}{\partial x} \right|^{-1} 
	= D(t_0,\,x_0^-;\,t)^{-1}	\,.
$$
Hence, the energy of the system at time~$t$ is 
\begin{equation}
\label{eq:energy_time}
E(t)	= \int_0^{a(t)} {\cal T}^{00} (t_0,x_0^-) \,
		D(t_0,\,x_0^-;\,t) \, \der x_0^-	\,.
\end{equation}

Note that since the local Doppler factor squared 
is the ratio of the energy densities 
at two consecutive reflection points, 
then it satisfies the following multiplicative property. 
Let $(t_1,x_1^-),\,(t_2,x_2^-),\,\ldots,\,(t_k,x_k^-)$ 
be space-time points on the characteristic connecting 
$(t_0, x_0^-)$ and $(t,x)$, such that at all of them 
the rays are going to the right, and let 
$t_0<t_1<\ldots<t_k<t$. Then
$$
D(t_0,\,x_0^-;\,t) = D(t_0,\,x_0^-;\,t_1)
	\, D(t_1,\,x_1^-;\,t_2)	\,\cdots
	\, D(t_{k-1},\,x_{k-1}^-;\,t_k)
	\, D(t_k,\,x_k^-;\,t)	\,.
$$
As can be seen from \Rf{eq:locDopfac}, 
these multiplicative properties are closely related 
to the chain rule for diffeomorphisms, 
\begin{equation}
\label{eq:chain_rule}
(G^n)'(\theta) = G'(G^{n-1}(\theta)) 
		\, G'(G^{n-2}(\theta)) \, \cdots
		\, G'(\theta)	\,. 
\end{equation}

The mathematical theory of dynamical systems 
contains many results about derivatives 
of highly iterated maps as above~\Rf{eq:chain_rule}. 
In Sec.~\ref{sec:Dop} we will be able 
to translate some of them into asymptotic 
properties of the field energy.

A simple and intuitively clear formula 
for the rate of change of the field energy 
can be obtained by using \Rf{eq:energy}, 
\Rf{eq:eq}, \Rf{eq:bc1}, \Rf{eq:bc2}, 
and integrating by parts:
\begin{eqnarray*}
E'(t)	&=& \displaystyle{\frac1{4\pi} \int_0^{a(t)} 
		(A_t A_{xx} + A_x A_{tx}) \, \der x
		+ \frac1{8\pi} a'(t) A_x(t,a(t))^2}	\\
	&=& \displaystyle{\frac1{4\pi} A_t(t,a(t)) A_x(t,a(t)) 
		+ \frac1{8\pi} a'(t) A_x(t,a(t))^2}	\\
	&=& \displaystyle{- \frac1{8\pi} a'(t) A_x(t,a(t))^2} \,.
\end{eqnarray*}
From this, we realize that the force 
experienced by the wall is 
$\frac1{8\pi} A_x(t,a(t))^2$.


\subsection{The inverse problem: determining 
	the mirror's motion given the circle map}
\label{sec:inverse}

It is important to know whether the notion 
of a ``typical'' $G$ is 
the same as the notion of 
a ``typical'' $a$ or a ``typical''~$F$ 
(in the mathematical literature people speak 
about ``generic'' maps, 
and in physical literature about 
``universal'' maps). 
We do not know the answer to this question, 
and here we will give some 
arguments showing that the answer 
is not obvious. 
In this paper we will not use 
``generic'' or ``universal''. 
Rather we will make explicit the non-degeneracy 
assumptions so that they can be checked 
in the concrete examples. 
In Sec.~\ref{sec:small_amplitude} 
we will show that some universal properties 
for families of circle maps do not apply 
to $G$ constructed according to \Rf{eq:Fexp} 
with $a(t) = \bar{a} + \varepsilon b(t)$.

While the function $a$ can be expressed in terms of $F$ as 
$a = \frac{F-\id}{2} \circ 
	\left(\frac{F+\id}{2}\right)^{-1}$, 
the relation between $G$ and $a$ is much harder to invert. 
We should have 
\begin{equation}
\label{eq:cond_a}
a(\theta) + a(G(\theta)) = \Gtil(\theta)	\,,
\end{equation}
where $\Gtil(\theta):=G(\theta)-\theta$, 
so for any $n$, 
$$
a(\theta) = \Gtil(\theta) - \Gtil(G(\theta)) + \cdots 
		+ (-1)^n \Gtil(G^n(\theta))	
		+ (-1)^{n+1} a(G^{n+1}(\theta))  \,.
$$
Hence, if $G^{2k}(\theta_0)=\theta_0\,(\mod 1)$, 
a necessary condition for the existence of $a$ is that 
\begin{equation}
\label{eq:cond_period}
\sum_{i=0}^{2k-1} (-1)^i \Gtil(G^i(\theta_0)) = 0	\,.
\end{equation}
An example of a $G$ where the above condition 
is not satisfied can be readily constructed. 
We note that if a map fails 
to satisfy~\Rf{eq:cond_period} and if 
$(G^{2k})'(\theta_0) \neq 1$, 
then all small perturbations will also 
fail to satisfy~\Rf{eq:cond_period}. 
Thus, there are whole neighborhoods of maps 
that cannot be realized as $G$ for a moving mirror.

On the other hand, given very simple $G$'s, 
it is easy to construct 
infinitely many $a$'s that satisfy \Rf{eq:cond_a} and that 
therefore lead to the same~$G$.
For example, for $G(\theta)=\theta+\frac12$, 
\Rf{eq:cond_a} amounts to 
$a(\theta+\frac12)+a(\theta)=\frac12$.
If we prescribe $a$ for $\theta$ in $[0,\frac12]$, 
then this equation determines $a$ on $[\frac12,1]$ 
(the only care needs to be exercised so that 
the two determinations of $a$ match at $\theta=\frac12$). 
A similar construction works when $G$ permutes 
several intervals -- if we prescribe $a(\theta)$ 
in an interval $I$, \Rf{eq:cond_a} 
determines $a(\theta)$ in $G(I)$.

In the case when $G$ is conjugate to an irrational 
rotation, $G=h^{-1}\circ R_\alpha \circ h$, 
then \Rf{eq:cond_a} is equivalent to 
$$
a\circ h^{-1} \circ R_\alpha + a\circ h^{-1} 
	= h^{-1} \circ R_\alpha - h^{-1}	\,.
$$
Then $a \circ h^{-1}$ can be determined 
using Fourier analysis, setting 
$h^{-1}(\theta) = \theta + \sum_{k=-\infty}^{\infty} 
	\widehat{\tau}_k e^{2\pi i k \theta}$, 
$a \circ h^{-1} (\theta) 
	= \theta + \sum_{k=-\infty}^{\infty} 
	\widehat{\psi}_k e^{2\pi i k \theta}$, 
which leads to 
\begin{equation}
\label{eq:homol}
\left(e^{2\pi i k \alpha} + 1 \right) \widehat{\psi}_k 
	= \left(e^{2\pi i k \alpha} - 1 \right) 
		\widehat{\tau}_k \,.
\end{equation}
Let us assume that 
$| k\alpha - n - \frac12 | 
	\geq {\rm const} \, |k|^{-\nu}$
for some $\nu\geq1$ 
(a condition of this type is called 
a Diophantine condition 
-- see definition~\ref{def:Dioph_cond}), 
and that $h^{-1}$ has $r$ derivatives 
(which implies that its Fourier coefficients 
$\widehat{\tau}_k$ satisfy 
$| \widehat{\tau}^k | \leq {\rm const}\, |k|^{-r}$). 
Then if $r>\nu+2$, then the coefficients 
$\widehat{\psi}_k$ define a smooth function 
(for more details see, e.g., \cite[Sec.\ XIII.4]{Herman1979}). 
Of course, once we know $a\circ h^{-1}$, 
then, since $h^{-1}$ depends only on $G$ 
and is therefore determined, we can obtain~$a$.

In summary, there are maps $G$ that 
do not come from any $a$ at all, 
come from infinitely many $a$'s, 
or come from one and only one $a$. 
The maps $F$ can always be obtained 
from one and only one~$a$.


\section{Maps of the circle}
\label{sec:maps-cir}

In this section we recall briefly
some facts from the theory  of the  dynamics of the 
orientation preserving homeomorphisms (OPHs) 
and diffeomorphisms (OPDs) of the 
circle $\SSS$, following closely 
the book of Katok and Hasselblatt 
\cite[Ch.\ 11, 12]{KatokHasselblatt1995}; 
see also \cite{deMelovanStrien1993}
and \cite{Herman1979}. 
This is a very rich theory and we will only 
recall the facts that we will need 
in the physical application.

We shall identify $\SSS$ with the quotient $\RR/\ZZ$ 
and use the universal covering projection
$$
\pi : \RR \to \SSS \equiv \RR/\ZZ : 
	x \mapsto \pi(x):= x \,(\mod 1) \,.
$$
Another way of thinking about 
$\SSS$ is identifying it with the 
unit circle in $\CC$, and using 
the universal covering projection 
$x\mapsto e^{2\pi i x}$.

Let $f:\SSS\to\SSS$ be an OPH and 
$F:\RR\to\RR$ be its {\em lift\/} 
to $\RR$, i.e., a map satisfying 
$f \circ \pi = \pi \circ F$. 
The fact that $f$ is an OPH implies that 
$F(x+1) = F(x) + 1$ for each $x\in \RR$, 
which is equivalent to saying that 
$F-\id$ is 1-periodic. 
The lift $F$ of $f$ is unique 
up to an additive integer constant.
If a point $x\in\SSS$ is $q$-periodic, i.e., $f^q(x)=x$, 
then $F^q(x)=x+p$ for some $p\in\NN$.


\subsection{Rotation number}
\label{sec:rot-num}

A very important number to associate 
to a map of the circle is its 
rotation number, introduced by Poincar\'{e}. 
It is a measure of the average amount 
of rotation of a point 
along an orbit.

\begin{defi}
Let $f:\SSS\to\SSS$ be an \OPH\ and $F:\RR\to\RR$ 
a lift of $f$. Define 
\begin{equation}
\label{eq:rotnum_def}
\tau_0(F) := \lim_{n\to\infty} 
		\frac{F^n(x)-x}{n} \,, \qquad 
\tau(f) := \tau_0(F) \,(\mod 1)  
\end{equation}
and call $\tau(f)$ a {\em rotation number} of $f$. 
\end{defi}

It was proven by Poincar\'{e} that the limit in 
\Rf{eq:rotnum_def} exists and is 
independent of $x$.  
Hence, $\tau(f)$ is well defined.

The rotation number is a very important 
tool in classifying 
the possible types of behavior 
of the iterates of the OPHs of $\SSS$. 
The simplest example of an OPH of 
$\SSS$ is the {\em  rotation\/} 
by $\alpha$ on $\SSS\equiv\RR/\ZZ$, 
$\fal:x\mapsto x+\alpha \,(\mod 1)$ 
(corresponding to a rotation by 
$2\pi\alpha$ radians on $\SSS$ 
thought of as the unit circle in $\CC$). 
The map $\Fal:x\mapsto x+\alpha$ is a lift of $\fal$, and 
$\tau(\fal)=\alpha \,(\mod 1)$. 
In the case of $r_\alpha$ there are two possibilities:
\begin{itemize}
\item[(a)] If $\tau(\fal) = p/q \in \QQ$, 
then $R_{p/q}^q(x) = x + p$ for each $x\in \RR$, 
so every point in $\SSS$ is $q$-periodic for $r_{p/q}$. 
If $p$ and $q$ are relatively prime, 
$q$ is the minimal period.
\item[(b)] If $\tau(\fal) \notin \QQ$, 
then $\fal$ has no periodic points; 
every point in $\SSS$ has a dense orbit. 
Thus, the $\alpha$- and $\omega$-limit sets of any point 
$x\in\SSS$ are the whole $\SSS$, 
which is usually described as saying that 
$\SSS$ is a {\it minimal set\/} for $\fal$. 
[Recall that $\alpha(x)$ is the set of the points 
at which the orbit of $x$ accumulates in the past, 
and $\omega(x)$ those points where it accumulates 
in the future.]
\end{itemize}


\subsection{Types of orbits of OPHs of the circle}
\label{sec:typ-orb}

To classify the possible orbits of OPHs of the circle, 
we need the following definition 
(for the particular case $f:\SSS\to\SSS$).

\begin{defi}
\label{def:hom_hetero}
\begin{itemize}
\item[(a)] 
On orbit ${\cal O}$ of $f$ is called {\em homoclinic\/} 
to an invariant set $T\in\SSS\setminus{\cal O}$ if 
$\alpha(x)=\omega(x)=T$ for any $x\in{\cal O}$. 

\item[(b)] An orbit ${\cal O}$ of $f$ is said to be {\em heteroclinic\/} 
to two disjoint invariant sets $T_1$ and $T_2$ if ${\cal O}$ is 
disjoint from each of them and $\alpha(x)=T_1$, $\omega(x)=T_2$ 
for any $x\in{\cal O}$.
\end{itemize}
\end{defi}

With this definition, the possible 
types of orbits of circle OPHs 
were classified by Poincar\'{e} 
\cite{Poincare1885} as follows 
(for a modern pedagogical treatment 
see, e.g., \cite[Sec.~11.2]{KatokHasselblatt1995}):
\begin{itemize}
\item[(1)] For $\tau(f)=p/q\in\QQ$, all orbits of $f$ are of 
		the following types:
	\begin{itemize}
	\item[(a)] a periodic orbit with 
	the same period as 
	the  rotation $r_{p/q}$ 
	and ordered in the same way as 
	an orbit of $r_{p/q}$;
	\item[(b)] an orbit homoclinic 
	to the periodic orbit 
	if there is only one periodic orbit; 
	\item[(c)] an orbit heteroclinic 
	to two different periodic 
	orbits if there are two or more periodic orbits.
	\end{itemize}
\item[(2)] When $\tau(f)\notin\QQ$, 
	the possible types of orbits are:
	\begin{itemize}
	\item[(a)] an orbit dense in $\SSS$ 
	that is ordered in the same 
	way as an orbit of $r_{\tau(f)}$ 
	(as are the two following cases); 
	\item[(b)] an orbit dense in a Cantor set;
	\item[(c)] an orbit homoclinic to a Cantor set.
	\end{itemize}
\end{itemize}

We also note that in cases 2(b) and 2(c), 
the Cantor set that has 
a dense orbit is unique and  
can be obtained as the set of 
accumulation points of any orbit.


\subsection{Poincar\'{e} and Denjoy theorems}
\label{sec:PoincareDenjoy}

Because of the simplicity of the  
rotations it is natural 
to ask whether a particular OPH of 
$\SSS$ is equivalent in some sense 
to a  rotation. To state the results, 
we give a precise definition 
of ``equivalence'' and the important 
concept of topological transitivity.

\begin{defi}
\label{def:conju}
\begin{itemize}
\item[(a)] Let $f:M\to M$ and 
$g:N\to N$ be $C^m$ maps, $m\geq0$.
The maps $f$ and $g$ are 
{\em topologically conjugate} if
there exists a homeomorphism 
$h:M\to N$ such that
$f=h^{-1}\circ g\circ h$.
\item[(b)] The map $g$ is 
a {\em topological factor} of $f$
(or $f$ is {\em semiconjugate} to $g$) if
there exists a surjective continuous 
map $h:M\to N$ such that
$h\circ f=g\circ h$; the map $h$ 
is called a {\em semiconjugacy}.   
\item[(c)] A map $f:M\to M$ is 
{\em topologically transitive}
provided the orbit, $\{f^k(x)\}_{k\in\ZZ}$, 
of some point $x$ is dense in $M$.
\end{itemize}
\end{defi}

The meaning of the conjugacy is that 
$g$ becomes $f$ under a change of variables, 
so that from the point of coordinate independent 
physical quantities, $f$ and $g$ are equivalent. 
The meaning of the semiconjugacy is that, 
embedded in the dynamics of~$f$, 
we can find the dynamics of~$g$. 

The following theorem of Poincar\'{e} 
\cite{Poincare1885} was chronologically 
the first theorem classifying circle maps. 

\begin{theorem}
[Poincar\'{e} Classification Theorem]
\label{poincTh}
Let $f:\SSS\to\SSS$ be an OPH with 
irrational rotation number. Then:
\begin{itemize}
\item[(a)] if $f$ is topologically transitive, 
then $f$ is topologically conjugate to the 
 rotation $r_{\tau(f)}$;
\item[(b)] if $f$ is not topologically 
transitive, then there exists
a non-invertible continuous monotone 
map $h:\SSS\to\SSS$ such that
$h \circ f = r_{\tau(f)} \circ h$; 
in other words, $f$ is semiconjugate 
to the  rotation $r_{\tau(f)}$.
\end{itemize}
\end{theorem}

If we restrict ourselves to considering not OPHs, but OPDs
of the circle, we can say more about the conjugacy problem.
An important result in this direction was the theorem
of Denjoy~\cite{Denjoy1932}.

\begin{theorem}[Denjoy Theorem]
\label{th:denjoy}
A $C^1$ OPD of $S^1$ with irrational 
rotation number and derivative
of bounded variation is topologically 
transitive and hence (according
to Poincar\'{e} theorem) topologically conjugate
to a  rotation. In particular, 
every $C^2$ OPD $f:\SSS\to\SSS$ 
is topologically conjugate to $r_{\tau(f)}$.
\end{theorem}

We note that this condition is extremely sharp. 
For every $\varepsilon>0$ there are 
$C^{2-\varepsilon}$ maps 
with irrational rotation number and 
semiconjugate but not conjugate 
to a  rotation (see~\cite[Sec.\ X.3.19]{Herman1979}).


\subsection{Smoothness of the conjugacy}
\label{sec:smoothness}

So far we have discussed only the conditions 
for existence of
a conjugacy $h$ to a  rotation, 
requiring $h$ to be only a
homeomorphism. Can anything more 
be said about the differentiability
properties of $h$ in the case 
of smooth or analytic maps of the circle? 
As we will see later, this is 
a physically important question 
since physical quantities such as 
energy density depend on 
the smoothness of the conjugacy.
To answer this question precisely, 
we need two definitions.

\begin{defi}
\label{def:Dioph_cond}
A number $\rho$ is called 
{\em Diophantine} of type $(K,\nu)$
(or simply of type~$\nu$) 
for $K>0$ and $\nu\geq1$, if 
$\left| \rho - \frac pq \right| > K \, |q|^{-1-\nu}$ 
for all $\frac pq \in \QQ$. 
The number $\rho$ is called 
{\em Diophantine} if it is Diophantine 
for some $K>0$ and $\nu\geq1$. 
A number which is not Diophantine 
is called a {\em Liouville number}.
\end{defi}

It can be proved that for $K\to0$, the set 
of all Diophantine numbers of type $(K,\nu)$ 
has Lebesgue measure as close to full as desired.

\begin{defi}
A function $f$ is said to be $C^{m-\delta}$ 
where $m\geq1$ is an integer
and $\delta\in(0,1)$, if it is $C^{m-1}$ 
and its $(m-1)$st derivative
is $(1-\delta)$-H\"{o}lder continuous, i.e., 
$$
\left|D^{m-1}f(x) - D^{m-1}f(y)\right|
	< {\rm const}\, \left|x-y\right|^{1-\delta}	\,.
$$
\end{defi}

The first theorem answering the question 
about the smoothness of the 
conjugacy was the theorem of Arnold \cite{Arnold1961}. 
He proved that if the analytic map 
$f:\SSS\to\SSS$ is sufficiently close 
(in the sup-norm) 
to a  rotation and $\tau(f)$ 
is Diophantine of type $\nu\geq1$, 
then $f$ is analytically conjugate 
to the  rotation $r_{\tau(f)}$, 
i.e., there exists an analytic function 
$h:\SSS\to\SSS$ such that 
$h\circ f=r_{\tau(f)}\circ h$. 
The iterative technique applied by Arnold 
was fruitfully used later in the proof 
of the celebrated Kol\-mo\-go\-rov-Ar\-nold-Mo\-ser 
(KAM) theorem -- see, e.g., \cite{Wayne1996}. 
Arnold's result was extended to the 
case of finite differentiability by Moser~\cite{Moser1966}. 
In such a case, the Diophantine exponent 
$\nu$ has to be related to the 
number of derivatives one assumes for the map.

Arnold's theorem is local, i.e., it 
is important that $f$ is close to a
rotation. Arnold conjectured that any analytic map 
with a rotation number in a set of full measure 
is analytically conjugate to a  rotation. 
Herman~\cite{Herman1979} proved that there exists a set 
${\cal A}\subset[0,1]$ of full Lebesgue measure 
such that if $f\in C^k$ for $3\leq k\leq\omega$ 
and $\tau(f)\in {\cal A}$, 
then the conjugacy is $C^{k-2-\varepsilon}$
for any $\varepsilon>0$. 
The set ${\cal A}$ is characterized in terms of
the growth of the partial quotients 
of the continued fraction expansions
of its members; all numbers in ${\cal A}$ 
are Diophantine of order $\nu$ 
for any $\nu\geq1$.
His result was improved by Yoccoz~\cite{Yoccoz1984} 
who showed that
if $f\in C^k$, $3\leq k\leq\omega$, 
$\tau(f)$ is a Diophantine number of type
$\nu\geq1$, and $k>2\nu-1$, then there exists a 
$C^{k-\nu-\varepsilon}$ conjugacy $h$ 
between $f$ and $r_{\tau(f)}$     
for any $\varepsilon>0$, and by several others.
The best result on smooth conjugacy we know of, 
is the following version of Herman's theorem 
as extended by Katznelson and 
Ornstein~\cite{KatznelsonOrnstein1989}.

\begin{theorem}[Herman, 
Katznelson and Ornstein] 
Assume that $f$ is a $C^k$ circle 
OPD whose rotation number 
is Diophantine of order $\nu$, and $k>\nu+1$. 
Then the homeomorphism $h$ which 
conjugates $f$ with the  
rotation $r_{\tau(f)}$ is of class 
$C^{k-\nu-\varepsilon}$ for any $\varepsilon>0$.
\end{theorem}

There are examples of $C^{2-\varepsilon}$ maps 
with a Diophantine rotation number arbitrarily close to 
a  rotation and not conjugated 
by an absolutely continuous 
function to a  rotation 
-- see, e.g.,~\cite{HawkinsSchmidt1982}.


\subsection{Devil's staircase, frequency locking, 
		Arnold's tongues}

Let $\{f_\alpha\}_{\alpha\in A}$ be 
a one-parameter family of circle 
OPHs such that $f_\alpha(x)$ is 
increasing in $\alpha$ for every $x$. 
Then the function $\alpha\mapsto\tau(f_\alpha)$ 
is non-decreasing.
(Since  the maps are only defined modulo an integer 
and so is the rotation number, some care needs 
to be taken  to define increasing and non-decreasing 
when some of the objects we are 
considering change integer parts.
What is meant precisely is that 
if one takes the numbers with their 
integer parts, they can be made 
increasing or non-decreasing. This is done in detail 
in  \cite[Sec.~11.1]{KatokHasselblatt1995}, 
and we will dispense with making it explicit 
since it does not lead to confusion.)

For such a family the following fact holds: 
if $\tau(f_\alpha)\notin\QQ$, 
then $\alpha\mapsto\tau(f_\alpha)$ 
is strictly increasing locally at $\alpha$; 
on the other hand, if $f_\alpha$ 
has rational rotation number 
and the periodic point is  attracting 
or repelling  (i.e., there is a neighborhood of 
the point that gets mapped into itself by forwards or
backwards iteration), then $\alpha\mapsto\tau(f_\alpha)$ 
is locally constant at this particular value of~$\alpha$, 
i.e., for all $\alpha'$ sufficiently close to~$\alpha$, 
$\tau(f_{\alpha'})=\tau(f_\alpha)$. 
The local constancy of the function 
$\alpha\mapsto\tau(f_\alpha)$ is known as 
{\em frequency (phase, mode) locking\/}. 
Note that, since the rotation number is
continuous, when it indeed changes, it 
has to go through rational numbers.
The described phenomenon suggests 
the following definition.

\begin{defi}
A monotone continuous function $\psi:[0,1]\to\RR$ 
is called a 
{\em devil's staircase} if there exists 
a family $\{I_\xi\}_{\xi\in\Xi}$ 
of disjoint open subintervals of $[0,1]$ 
with dense union such that 
$\psi$ takes constant values on these subintervals. 
(We call attention to the fact that the 
complement of the intervals 
in which the function is constant can 
be of positive measure.)

The devil's staircase is said to be {\em complete\/} 
if the union of all intervals $I_\xi$ 
has a full Lebesgue measure.
\end{defi}

A very common way of phase locking 
for differentiable mappings 
arises when the map we consider 
has a  periodic point and that 
the derivative of the return map 
at the periodic point is not equal to~$1$. 
By the implicit function theorem, 
such a periodic orbit persists, 
and the existence of a periodic orbit implies that 
the rotation number is locally constant. 
At the end of the phase locking interval 
the map has derivative one and 
experiences a saddle-node (tangent) bifurcation.

We note that, unless certain combinations 
of derivatives vanish (see, e.g., \cite{Ruelle1989}), 
the saddle-node bifurcation 
happens in a universal way. 
That is, there are analytic changes of variables 
sending one into another. 
This leads to quantitative predictions. 
For example, the Lyapunov exponents 
of a periodic orbit should behave as a square root 
of the distance of the parameter to the edge 
of the phase locking interval.

Of course, other things can happen in special cases: 
the fixed point may be attractive but only neutrally so, 
there may be an interval of fixed points, 
the family may be such that there are no 
frequency locking intervals (e.g., the  rotation).
Nevertheless, all these conditions are exceptional
and can be excluded in concrete examples 
by explicit calculations. (For example, if 
the family of maps is analytic but not a root of 
the identity,
it is impossible to have an interval of fixed points.)

In the example we will consider, we will not perform 
a complete proof that a devil's staircase occurs, 
but rather we will present numerical evidence. 
In particular, the square root behavior 
of the Lyapunov exponent with the distance 
to the edge of the phase locking interval 
seems to be verified.

Let us now consider two-parameter 
families of OPDs of the circle, 
$\{\phi_{\alpha,\beta}\}$, depending 
smoothly on $\alpha$ and $\beta$. 
Assume that when $\beta=0$, the maps of the family 
are  rotations by $\alpha$, 
i.e., $\phi_{\alpha,\,0}=r_\alpha$. 
We will call $\beta$ the {\em nonlinearity parameter}. 
Assume also that 
$\partial\phi_{\alpha,\beta}/\partial\alpha>0$. 
An example of this type is the family 
studied by Arnold~\cite{Arnold1961}, 
\bequ{eq:Arnmap}
\eta_{\alpha,\beta} : \SSS \to \SSS : 
	x \mapsto \eta_{\alpha,\beta} (x) 
	:= x + \alpha + \beta\sin 2\pi x  \,(\mod 1)	\,,
\eequ
where $\alpha\in [0,1)$, $\beta\in(0,1/2\pi)$.

The rotation number $\tau$ is a continuous 
map in the uniform topology, 
and $\phi_{\alpha,\beta}$ is a continuous 
function of $\alpha$ and $\beta$, so the function 
$(\alpha,\beta)\mapsto\tau(\phi_{\alpha,\beta})
	=:\tau_\beta(\alpha)$  
depends continuously on $\alpha$ and $\beta$. 
The map $\tau_\beta$ is non-decreasing; 
for $\beta>0$, $\tau_\beta$ is locally constant 
at each $\alpha$ for which $\tau_\beta(\alpha)$ 
is rational 
and strictly increasing if $\tau_\beta(\alpha)$ 
is irrational. 
Thus, $\tau_\beta$ is a devil's staircase.

Since $\tau_\beta$ is strictly increasing 
for irrational values of $\tau_\beta(\alpha)$, the set 
$I_\nu := \{(\alpha,\beta)\,|\,\tau_\beta(\alpha) = \nu\}$
for an irrational $\nu\in[0,1]$ is a graph 
of a continuous function. 
For a rational $\nu$, $I_\nu$ has 
a non-empty interior and is 
bounded by two continuous curves. 
The wedges between these two curves 
are often referred to as {\em Arnold's tongues}.

The fact that 
$\tau(\phi_{\alpha,0})=\tau(r_\alpha)=\alpha$ implies 
that for $\beta=0$, the set of $\alpha$'s for which 
there is frequency locking coincides 
with the rational numbers between~0 
and~1, so its Lebesgue measure is zero. 
When $\beta>0$, its Lebesgue measure is positive. 
The width of the Arnold's tongues for small $\beta$ 
for the Arnold's map \Rf{eq:Arnmap} 
is investigated, e.g., in \cite{Davie1996}.
Much of this analysis carries out 
for more general functions 
such as the ones we encounter 
in the problem of the periodically 
pulsating resonator.

The total Lebesgue measure of the frequency locking intervals, 
$m( \{\tau^{-1}_\beta (\nu) \,| \, \nu\in \QQ\cap[0,1] \} )$, 
becomes equal to~$1$ when the family 
of circle maps consists of maps 
with a horizontal point 
(so  that the map, even if having a continuous 
inverse, fails to have a differentiable one) 
-- see 
\cite{Jensenetal1984,Lanford1985} for numerical results 
and \cite{Swiatek1988} for analytical proof. With the 
Arnold's map $\eta_{\alpha,\beta}$ this 
happens when $\beta=1/2\pi$.
In our case this happens when the mirror 
goes at one instant at the 
speed of light.

We note also that the numerical papers 
\cite{Shenker1982,Jensenetal1984,%
Lanford1985,Cvitanovicetal1985} 
contain not only conjectures about the measure
of the phase locking intervals but, perhaps 
more importantly, conjectures about scaling relations 
that hold ``universally''. 
In particular, the dimension of the set 
of parameters not covered 
by the phase locking intervals should be 
the same for all non-degenerate families. 
These universality conjectures are supported 
not only by numerical evidence but also 
by a  renormalization group picture 
-- see, e.g., \cite{Lanford1986} 
and the references therein. 
These universality predictions have been 
verified in several 
physical contexts. Notably in turbulence
by Glazier and Libchaber
~\cite{GlazierLibchaber1988}.

As we will see in Sec.~\ref{sec:small_amplitude}, 
we do not expect that the 
families obtained in (\ref{eq:Fexp}) 
for mirrors oscillating with different amplitudes
belong to the same universality class as
typical mappings, but they should have 
universality properties that are 
easy to figure out from those of 
the above references.


\subsection{Distribution of orbits}
\label{sec:distr}

For the physical problem at hand it is also important 
to know how the iterates of the circle map 
$x\mapsto g(x):=G(x)(\mod 1)$ are distributed. 
As we shall see in lemma~\ref{cor:measure}, 
if the iterates of $g$ are well distributed 
(in an appropriate sense), 
the energy of the field in the resonator 
does not build up.
The distribution of an orbit is conveniently 
formalized by using the concept of invariant measures. 
We recall that a measure $\mu$ on $X$ is {\em invariant\/} 
under the measurable map $f:X\to X$ if 
$\mu(f^{-1}(A)) = \mu(A)$ for each measurable set $A$.

Given a point $x\in\SSS$, the frequency of visit of 
the orbit of $x$ to $I\subset\SSS$ can be defined by
\begin{equation}
\label{eq:inv_measure}
\mu_x(I) := \lim_{n\to\infty} \frac{\# \{ i\,|\,0\leq i\leq n 
	\mbox{  and  } f^i(x)\in I \} }{n}	\,.
\end{equation}
It is easy to check that if for every interval $I$, 
the limit \Rf{eq:inv_measure} exists, 
it defines an invariant  measure describing 
the frequency of visit of the orbit of $x$.
Therefore, if there are orbits which 
have asymptotic frequencies of visit, 
we can find invariant measures.

A trivial example of the existence of 
such measures is when $x$ is 
periodic. In such a case,  the measure
$\mu_x$ is a sum of  Dirac delta  functions 
concentrated on the periodic orbit.
The measure of an interval is proportional 
to the number of points in the orbit it contains.
We also note that it is easy to construct 
two-dimensional systems 
[e.g., $(r,\theta) \mapsto 
	(1 + 0.1 (r-1), \, \theta + (r-1)^2 \sin(\theta)^2)$] 
for which the limits like the one in \Rf{eq:inv_measure} 
do not exist except for measures concentrated on 
the fixed points, so that even the existence of 
such equidistributed orbits is not obvious.

There are also relations going in the opposite 
direction -- if invariant measures exist, they 
imply the existence of well distributed orbits. 
We recall that the 
Krylov-Bogolyubov theorem 
\cite[Thm.~4.1.1]{KatokHasselblatt1995} 
asserts that any continuous map on 
a compact metrizable space has an 
invariant probability measure. 
Moreover, the Birkhoff ergodic theorem 
\cite[Thm.~4.1.2]{KatokHasselblatt1995} 
implies that  given any invariant measure
$\mu$, the set of points for 
which $\mu_x$ as in \Rf{eq:inv_measure}
does not exist has measure zero.

Certain measures have the property 
that $\mu_x = \mu$ for $\mu$-almost all 
points. These measures are called 
{\em ergodic}. (There are several 
equivalent definitions of ergodicity
and this is one of them.)
From the physical point of view, 
we note that a measure is 
ergodic if all the points in 
the measure are distributed according to it. 
For maps of the circle,
there are several criteria that allow to conclude that a
map is ergodic.

For rotations of the circle 
with an irrational rotation number
we recall the classical 
Kronecker-Weyl equidistribution theorem 
\cite[Thm.~4.2.1]{KatokHasselblatt1995} 
which  shows that any irrational rotation 
is uniquely ergodic, i.e., 
has only one invariant measure -- 
the Lebesgue measure~$m$. 
(Such uniquely ergodic maps are, 
obviously, ergodic because, 
by Birkhoff ergodic theorem, the 
limiting distribution has to exist
almost everywhere, but, since there is 
only one invariant measure, all these
invariant distributions have to 
agree with the original measure.) 
Thus, the iterates of any $x\in\SSS$ under 
an irrational  rotation 
are uniformly distributed on the circle.

For general non-linear circle OPDs 
the situation may be quite different. 
As an example, consider Arnold's map 
$\eta_{\alpha,\beta}$~\Rf{eq:Arnmap}. 
If it is conjugate to an irrational rotation by $h$, 
i.e., $\eta_{\alpha,\beta} = h^{-1} \circ 
	r_{\tau(\eta_{\alpha,\beta})} \circ h$, 
then there is a unique invariant 
probability measure $\mu$ defined 
for each measurable set $A$ by
$\mu(A) := m(h(A))$. 
This implies that if $I$ is an interval 
in $\SSS$, then the frequency 
with which a point $x$ visits $I$ 
is equal to $\mu(I)$. 

\begin{figure}
\centerline{\epsfig{file=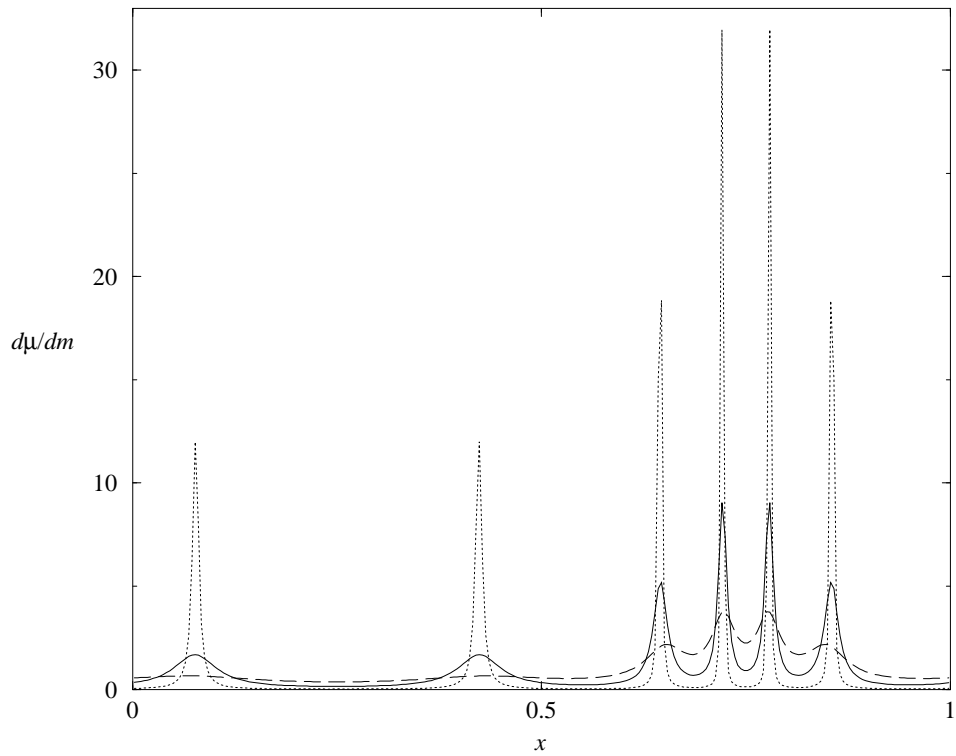, height= 5 in, angle=-90}} 
\bigskip
\caption{Density of the invariant measures 
for $\beta=0.1$ and 
$\alpha=0.253$ (dashed line), 
$\alpha=0.2539$ (solid line), 
and $\alpha=0.253975$ (dotted line).} 
\label{fig:singular_measures}
\end{figure}

On the other hand, if 
$\tau(\eta_{\alpha,\beta}) = p/q \in \QQ$, then 
all orbits are periodic or asymptotic to periodic. 
Thus, the only possible invariant measure 
is concentrated at the periodic points 
and therefore singular, 
if the periodic points are isolated. 
Let us now assume that $\alpha$ is very close to 
$\tau_\beta^{-1}(p/q)$, but does not belong to it. 
Then $\eta_{\alpha,\beta}$ has no periodic orbits, 
but still there exists 
a point $x$ which is ``almost periodic'', 
i.e., the orbits linger 
for an extremely long time near the points 
$x,\, \eta_{\alpha,\beta}(x),\, \cdots ,\,
		\eta_{\alpha,\beta}^{q-1}(x)$. 
So that, even if the invariant measure 
is absolutely continuous, 
one expects that it is nevertheless quite peaked 
around the periodic orbit 
-- see Fig.~\ref{fig:singular_measures}.
The behavior of such maps is described quantitatively 
by the ``intermittency theory'' 
\cite{PomeauManneville1980}.

The continuity properties of the measures 
of the circle are not so easy to ascertain. 
Nevertheless, there are certain results that are 
easy to establish. 

Of course, in 
the case that we have a rational 
rotation number and 
isolated periodic orbits, some 
of them attracting and some of them 
repelling, the only possible 
invariant measures are 
measures concentrated in the 
periodic orbits.

For the irrational rotation number case, 
the Kronecker-Weyl theorem implies that all the maps 
with an irrational rotation number
-- since they are semi-conjugate to a  rotation 
by Poincar\'{e} theorem -- are uniquely ergodic. 
In the situations where Herman's theorem applies, 
this measure will have a smooth density 
since it is the push-forward of Lebesgue measure 
by a smooth diffeomorphism.

We also recall that by Banach-Alaoglu theorem 
and the Riesz representation theorem, 
the set of Borel probability measures is compact  
when we give it the topology of
$\mu_n \rightarrow \mu \iff \mu_n(A) \rightarrow \mu(A) $ 
for all Borel measurable sets~$A$.
(This convergence is called 
weak-* convergence by functional analysts 
and convergence in probability by probabilists.)

\begin{lem}
\label{continuity}
If $\lambda^*$ is a parameter 
value for which 
$f_{\lambda^*} $ admits only
one invariant measure
$\mu_{\lambda^*}$,
given
$\mu_{\lambda_i} $ invariant measures
for $f_{\lambda_i}$, 
with $\lambda_i \rightarrow \lambda^*$,
then
$\mu_i$ converges  in the weak-* sense to 
$\mu_{\lambda^*}$.
\end{lem}

Note that we are not assuming that
$f_{\lambda_i}$ are uniquely ergodic. 
In particular, the lemma 
says that in the set of uniquely 
ergodic maps, the  map that a parameter 
associates the invariant measure is 
continuous if we give the measures 
the topology of weak-* convergence.

\proofend

\noindent
{\bf Proof.} 
Let $\mu_{\lambda_{i_k}}$ be a 
convergent subsequence.
The limit should be an invariant
measure for  $f_{\lambda^*}$.
Hence, it should be $\mu_{\lambda^*}$. 
It is an easy point set topology lemma
that for functions taking values in a 
compact metrizable space, if all subsequences converge
to the same point, then this point is a limit. 
The space of measures with weak-* topology 
is metrizable because by Riesz representation theorem 
is the dual of the space of continuous functions with sup-norm, 
which is metrizable.

\proofend

We also point out that as a corollary of 
KAM theory \cite{Arnold1961} we can obtain that 
for non-degenerate families, 
if we consider the parameter 
values for which the rotation number is
Diophantine with uniform constants, 
the measures are differentiable
jointly on $x$ and in the parameter. 
(For the differentiability in the 
parameter, we  need to 
use Whitney differentiability or,
equivalently, declare that there is
a family of densities
differentiable both in $x$ and 
in $\lambda$ that agrees 
with the densities for these values of 
$\lambda$.)

On the other hand, we point out that there are situations 
where the invariant  measure is not unique 
(e.g., a rational rotation or a map 
with more than one periodic orbit). 
In such cases, it is not difficult to 
approximate them by maps in such a way that the invariant 
measure is discontinuous in the weak-* topology as 
a function of the parameter.
The discontinuity of the measures 
with respect to parameters, as we shall see, 
has the physical interpretation that, by changing 
the oscillation parameters by arbitrarily small amounts, 
we can go from unbounded growth in the energy 
to the energy remaining bounded.


\section{Application to the resonator problem}
\label{sec:appl-th}

Now we return to the problem of a 
one-dimensional optical resonator 
with a periodically moving wall to discuss 
the physical implications of circle maps theory, 
and illustrate with numerical results in an example.


\subsection{Circle maps in the resonator problem}

If we take $a(t)$ to depend on two parameters, 
$\alpha$ and $\beta$, as in \Rf{equ:a1}, 
then, as we saw in Sec.~\ref{sec:sol_problem}, 
the time between the consecutive reflections 
at the mirrors can be described in terms 
of the functions $F_{\alpha,\beta}$ 
and $G_{\alpha,\beta}$ defined by \Rf{eq:Fexp}. 
These maps are lifts of circle maps 
that we will denote by 
$f_{\alpha,\beta}$ and $g_{\alpha,\beta}$. 
The restriction on the range of $\beta$ in 
\Rf{equ:a1} implies that $f_{\alpha,\beta}$ 
and $g_{\alpha,\beta}$ are analytic circle OPDs. 
Therefore, we can apply the results 
about the types of orbits 
of OPHs of $\SSS$, Poincar\'{e} and Denjoy theorems, 
as well as the smooth conjugacy results 
and the facts about the distribution of orbits.

In an application where the motion 
of the mirror [i.e., $a(t)$] 
is given, one needs to compute 
$F_{\alpha,\beta}$ and 
$G_{\alpha,\beta}$~\Rf{eq:Fexp}, 
which cannot be expressed explicitly 
from $a(t)$ but they require only to solve 
one variable implicit equation. 
In the numerical computations 
we used the subroutine {\sc zeroin} \cite{Forsytheetal1977} 
to solve implicit equations. 
If $y=F_{\alpha,\beta}(t)$ and $z=G_{\alpha,\beta}(t)$, 
then for $a(t)$ given by \Rf{equ:a1}, 
$y$ and $z$ are given implicitly by
\barrz
&- y + t + \alpha + 2\beta \sin[\pi(y+t)] = 0	\\
&- z + t + \alpha + \beta \left[\sin(2\pi t) + 
		\sin(2\pi z) \right] = 0	\,.
\earrz
Given $t$, we can find $y$, $z$ applying {\sc zeroin}.


\subsection{Rotation number, phase locking}
\label{sec:rot_num}

In this section, our goal is to translate the 
mathematical predictions from the theory of 
circle maps into physical predictions for the
resonator problem.

\begin{figure}
\centerline{\epsfig{file=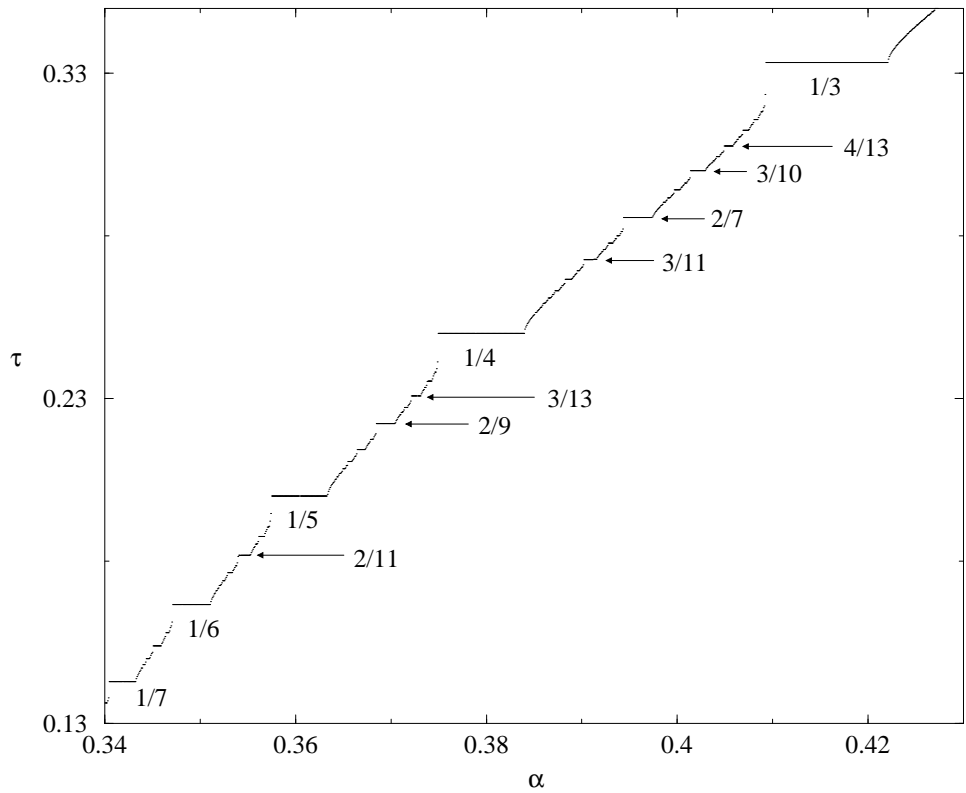, height= 5 in, angle=-90}} 
\bigskip
\caption{A part of the graph of 
	$\tau(g_{\alpha,1/2\pi})$ vs.\ $\alpha$.}
\label{fig:rotation_numbers}
\end{figure}

The theory of circle maps guarantees that 
the  measure of the  frequency
locking intervals for $g_{\alpha,\beta}$
is small when $\beta$ is small and 
becomes $1$ when
$\beta=1/2\pi$. The theory also guarantees
for analytic maps
that, unless a power of the map 
is the identity, 
the frequency locking intervals are non-trivial.
For the example that we have at hand, it is very easy 
to verify that this does not happen and, therefore, 
we can predict that there will be 
frequency locking intervals and that
as the amplitude 
of the oscillations of the moving mirror
increases so that 
the maximum speed of the moving mirror reaches the 
speed of light, the devil's staircase becomes complete.
Fig.~\ref{fig:rotation_numbers} 
%
%
shows a part of the complete devil's staircase -- 
the situation which happens when 
the maps $g_{\alpha,\beta}$ 
and $f_{\alpha,\beta}$ lose their invertibility, 
i.e., for $\beta=1/2\pi$.

We also recall that the theory 
of circle maps makes predictions 
about what happens for 
non-degenerate phase locking 
intervals. Namely, for parameters inside 
the phase locking interval the map 
has a periodic fixed point 
and the Lyapunov exponent is smaller than~$0$, 
while at the edges of the phase locking interval 
the map experiences a non-degenerate 
saddle-node bifurcation
-- provided that certain combinations 
of the derivatives do not vanish~\cite{Ruelle1989}.

\begin{figure}
\centerline{\epsfig{file=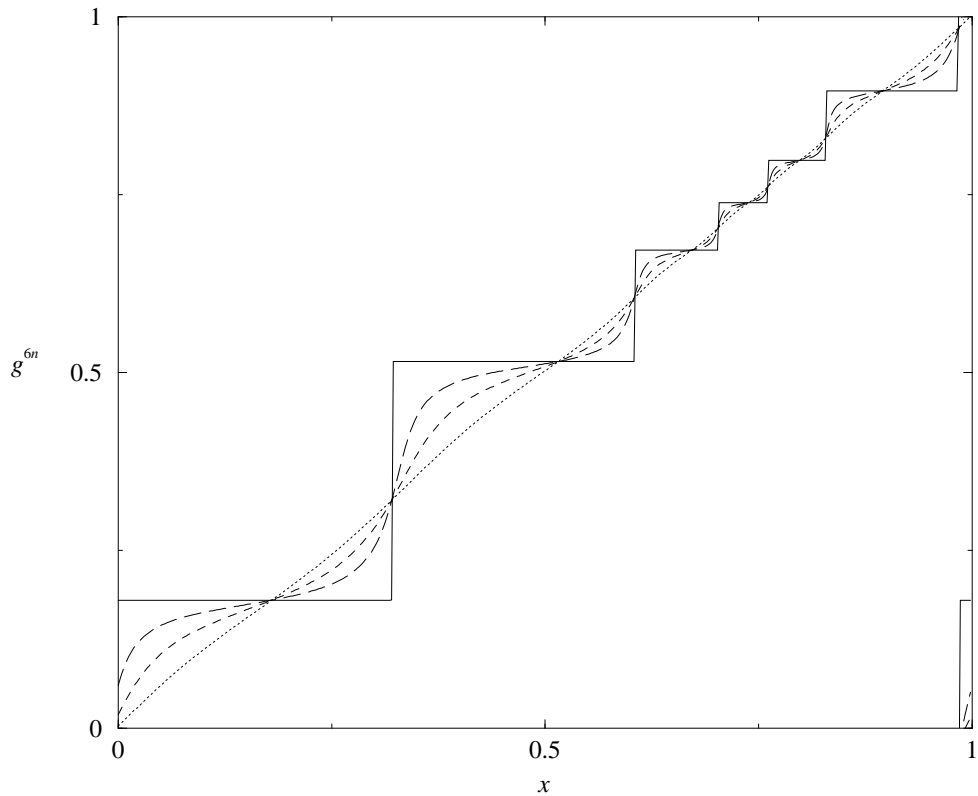, height= 5 in, angle=-90}} 
\bigskip
\caption{Development of the 
			piecewise-constant structure 
			of $g^{6n}_{0.2545,\,0.1}$ 
			(the rotation number of 
			$g_{0.2545,\,0.1}$ 
			is~$1/6$). 
			Graphs of $g^{6n}_{0.2545,\,0.1}$
			are plotted for 
			$n=1$ (dotted line), 
			$n=5$ (dashed line), 
			$n=10$ (long dashed line), 
			$n=100$ (solid line).} 
\label{fig:staircase_for_g}
\end{figure}

We note that for parameters for which the map
is in non-degenerate frequency locking, 
i.e., $\tau(g_{\alpha,\beta}) = p/q$ 
and the attractive periodic point of period $q$ 
has a negative Lyapunov exponent, 
$\{G^{nq}_{\alpha,\beta}(x)\}_{n=0}^\infty$ 
will converge exponentially to the fixed point for all 
$x$ in a certain interval, according to the results 
about the types of orbits of 
circle maps (Sec.~\ref{sec:typ-orb}). 
The whole circle
can be divided into such intervals and 
a finite number of periodic points. 
Therefore, the graph of
$G^{nq}_{\alpha,\beta}$, and hence of 
$g^{nq}_{\alpha,\beta}$, 
will look -- up to errors exponentially 
small in~$n$ -- like a piecewise-constant function 
%
%
with values (up to integers) in the fixed points of 
$g^{q}_{\alpha,\beta}$ -- see Fig.~\ref{fig:staircase_for_g}. 
The fact that certain functions tend to piecewise-constant functions 
for large values of the argument (which follows from what we found about 
$G^{nq}_{\alpha,\beta}$ for large $n$) was observed numerically 
for particular motions of the mirror 
in \cite{Law1994a,ColeSchieve1995}.
In physical terms, this means that the rays will be 
getting closer and closer together, 
so with the time the wave packets 
will become narrower and narrower and 
more and more sharply peaked. 
The number of wave packets is equal to $q$.
The number of reflections from the moving mirror per unit time 
will tend to the inverse of the rotation number.
In the next section we discuss how this yields 
an increase of the field energy which happens 
exponentially fast on time.

The fact that for $\tau(g_{\alpha,\beta}) \in\QQ$ 
the rays approach periodic orbits, is also interesting from 
a quantum mechanical point of view due to the relation between 
the periodic orbits in a classical system and 
the energy levels of 
the corresponding quantum system, given 
by the Gutzwiller's trace formula
(see, e.g., \cite{Gutzwiller1990}).

We also note that we expect that 
slightly away from the edges of 
a phase locking interval, the invariant density 
will be sharply peaked around the points in which 
it was concentrated in the phase locking intervals.
This is described by the ``intermittency theory'' 
~\cite{PomeauManneville1980}.

To observe numerically in 
our example what happens when $\alpha$ enters or 
leaves a frequency locking interval, we set 
$\calN{\beta}{\nu} := \{\alpha\in[0,1) \, | \, 
	\tau(g_{\alpha,\beta}) = \nu \}$. 
Fig.~\ref{fig:singular_measures} 
represents the Radon-Nikodym 
derivative ${\der\mu}/{\der m}$ of the invariant probability 
measure $\mu$ with respect to the Lebesgue measure $m$ 
[i.e., of the density of the invariant measures, 
which, as we saw in \Rf{eq:inv_measure}, 
is the frequency of visit of the iterates]. 
%
%
The figure shows ${\der\mu}/{\der m}$ for 
$\alpha$ close to the left end of $\calN{0.1}{1/6}$. 
When $\alpha$ approaches (from the left) 
the left end of $\calN{0.1}{1/6}$, 
${\der\mu}/{\der m}$ becomes sharply peaked at some points, 
and when $\alpha$ enters the frequency locking interval, 
the invariant measure becomes singular 
($g_{\alpha,\,0.1}$ undergoes tangent bifurcation 
at $\alpha = 0.253977\ldots$). 
All seems to be consistent 
with the  conjecture that all the frequency-locking intervals
in the family (away of $\beta = 0$) are non-degenerate, 
i.e., that at the boundaries of the phase locking intervals 
the map satisfies the hypothesis 
of the saddle-node bifurcation theorem.


\subsection{Doppler shift}
\label{sec:Dop}

One of the most interesting parts of the applications 
of circle map theory is the ease with which we can describe 
the effect on the energy after repeated reflections.

Recall that in Sec.~\ref{sec:Dop_shift}, we found the time 
dependence of the field energy under the assumption 
that at time $t$ all rays are going to the right. 
This assumption is not very restrictive in the case of 
a rational rotation number since, as we found in 
Sec.~\ref{sec:rot_num}, the field develops wave packets that 
become narrower with time, so  
\Rf{eq:locDopfac} and \Rf{eq:energy_time} 
hold for the asymptotic behavior of the energy. 
Note that \Rf{eq:locDopfac} expresses the Doppler shift 
factor in terms of the derivatives of the map~$G$. 
This gives a very close relation between the dynamics 
and the behavior of the wave packets.

\begin{prop}
Let $\alpha$ and $\beta$ be such that 
$\tau(g_{\alpha,\beta})=p/q$, and 
that the map $G:=G_{\alpha,\beta}$ 
has a stable periodic orbit 
$\Theta_q=\{\theta_1,\,\ldots,\,\theta_q\}$ 
such that $(G^q)'(\theta_1)<1$. 
Assume that the initial 
electromagnetic field in the cavity 
is not zero at some space-time point 
for which the phase of the first reflection 
from the moving mirror 
is in the basin of attraction of~$\Theta_q$. 

Then the energy of the field in the resonator 
will be asymptotically increasing 
at an exponential rate: 
\begin{equation}
\label{eq:energy_exp}
E(t) \sim \exp \left\{\frac{\ln D(\Theta_q)}{p}\,t\right\} \,.
\end{equation}
\end{prop}

\noindent
{\bf Proof.}
First notice that 
the number of reflections from the moving mirror 
per unit time reaches a well defined limit 
(one and the same for all rays) 
-- the inverse of the rotation number. 
Secondly, as was discussed in Sec.~\ref{phys_set}, 
at reflection from the moving mirror at phase~$\theta$, 
a wave packet becomes narrower by a factor 
of $D(\theta)$~(\ref{Doppler_factor}), 
which leads to a $D(\theta)$~times increase in its energy. 
Asymptotically, the phases at reflection 
will approach the stable periodic orbit 
$\Theta_q=\{\theta_1,\,\ldots,\,\theta_q\}$
of $g_{\alpha,\beta}$. The Doppler factors 
at reflection will tend correspondingly to 
$\{D(\theta_1),\,\ldots,\,D(\theta_q)\}$ ~(\ref{Doppler_factor}). 
Hence, in time~$p$ each ray will undergo $q$ reflections 
from the moving mirror, the total Doppler shift factor 
along the periodic orbit $\Theta_q$ being 
$$
D(\Theta_q) := \prod_{i=1}^{q} D(\theta_i)
	= \prod_{i=1}^{q} \displaystyle{
		\frac{1-a'(\theta_i)}{1+a'(\theta_i)}}	\,.
$$

On the other hand, the definition of the map $G$ as 
the advance in the time between successive reflections 
from the moving mirror yields 
$\theta_i=G^{i-1}(\theta_1)$. 
The chain rule applied to the explicit 
expression \Rf{eq:Fexp} for $G$ yields 
$$
(G^{q-1})' (\theta_1) 
	= \prod_{j=1}^{q-1} G'(\theta_j) 
	= \prod_{j=1}^{q-1} \displaystyle{
		\frac{1+a'(\theta_j)}{1-a'(\theta_{j+1})}} \,,
$$
which gives the following expression for $D(\Theta_q)$ 
[cf.~\Rf{eq:locDopfac}]: 
\begin{equation}
\label{eq:Dop_orbit}
D(\Theta_q) := \displaystyle{
		\frac{1-a'(\theta_1)}{1+a'(\theta_q)} 
		\left[ (G^{q-1})' (\theta_1) \right]^{-1}
	= \frac{1-a'(\theta_1)}{1+a'(\theta_q)} 
		(G^{1-q})' (\theta_q) }		\,.
\end{equation}
Hence, the energy density grows by a factor of $D(\Theta_q)^2$. 
Since after $q$ reflections the wave packet is concentrated 
in a length $D(\Theta_q)$ times smaller, the total energy 
grows by a factor of $D(\Theta_q)$ in $p$ units of time, 
which implies~\Rf{eq:energy_exp}.

\proofend

The quantities $(G^n)'(\theta)$ that appear 
in \Rf{eq:Dop_orbit} have been studied intensively 
in dynamical systems theory since they control the growth 
of infinitesimal perturbations of trajectories. 
Similarly, they are factors that multiply 
the invariant densities 
when they get transported, as we will see 
in~\Rf{eq:inv_density}.

\begin{figure}
\centerline{\epsfig{file=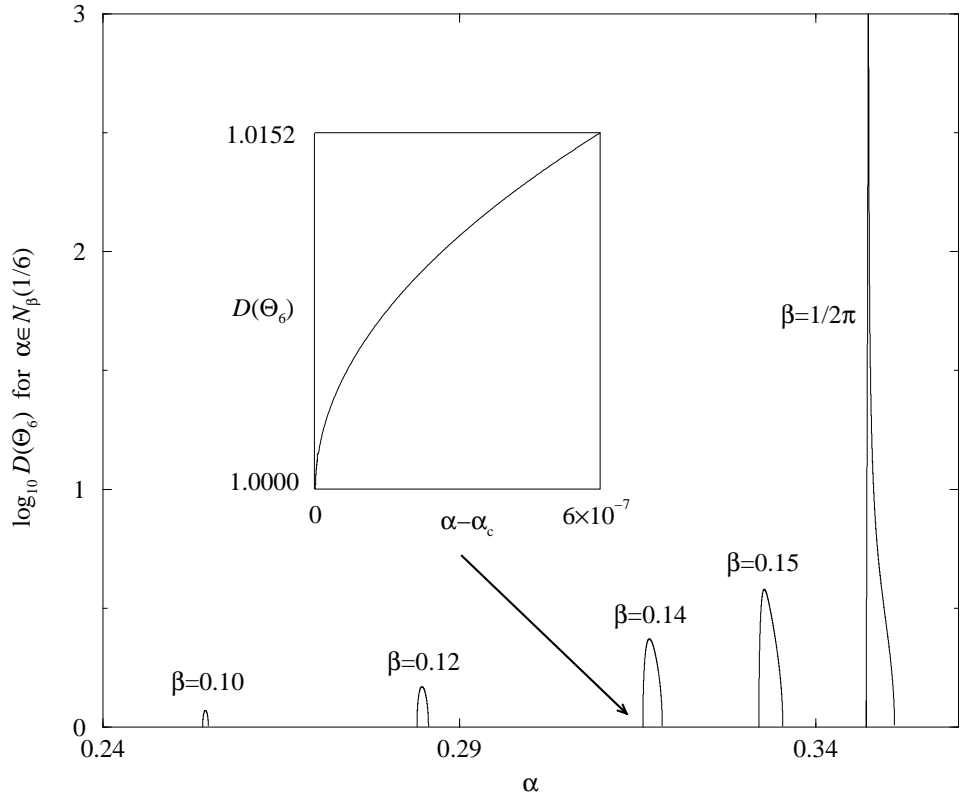, height= 5 in, angle=-90}} 
\bigskip
\caption{A log-linear graph of the total Doppler factor 
		for $g_{\alpha,\beta}$ 
		in the phase locking interval 
		of rotation number $1/6$ 
		for different $\beta$. 
		The insert [linear-linear graph 
		of $D(\Theta_6)$ 
		vs.\ $\alpha-\alpha_{\rm c}$] 
		calls attention to the square-root behavior 
		at edges; $\alpha_{\rm c}$ is the value of 
		$\alpha$ at the left end of $\calN{0.14}{1/6}$.}
\label{fig:log_doppler}
\end{figure}

We found numerically the total Doppler factors $D(\Theta_q)$ 
for some particular choices of the parameters. 
%
In Fig.~\ref{fig:log_doppler}, 
$\log_{10} D(\Theta_6)$ 
is shown for different values of $\beta$ and 
for $\alpha\in\calN{\beta}{1/6}$. 
Obviously, the maximum value of $D(\Theta_6)$ 
depends strongly on~$\beta$, becoming infinite 
for $\beta=1/{2\pi}$ and some $\alpha\in N_{1/2\pi}(1/6)$. 
For smaller values of~$\beta$, 
the Doppler factor is much smaller. 
Moreover, the width of the frequency locking intervals 
for small~$\beta$ is small, 
so the probability of hitting a frequency 
locking interval with arbitrarily 
chosen $\alpha$ and $\beta$ is small. 
[The likelihood of frequency locking 
for the Arnold's map~(\ref{eq:Arnmap}) 
is studied numerically in~\cite{Lanford1985}.]

In the case when Herman's theorem apply, 
the derivatives of $G^n$ are bounded independently of $n$, 
which causes the energy of the system 
to be bounded for all times, which is proved 
in the following proposition.

\begin{prop}
If $G_{\alpha,\beta}$ is such that it satisfies 
the hypothesis of Herman's theorem, 
then the energy density remains bounded for all times.
\end{prop}

\proofend

\noindent
{\bf Proof.} 
In such a case $G_{\alpha,\beta}=h^{-1}\circ R\circ h$ 
with $h$ differentiable and $R$ a rotation 
by $\tau(g_{\alpha,\beta})$. Therefore 
$G_{\alpha,\beta}^n = h^{-1} \circ R^n \circ h$ 
and 
$$
(G_{\alpha,\beta}^n)' (\theta) = (h^{-1})' (R^n\circ h(\theta)) 
		\, (R^n)' (h(\theta)) \, h'(\theta)
	= (h^{-1})' (R^n\circ h(\theta)) \, h'(\theta)
$$
because $(R^n)'=1$. The two factors in the right-hand side 
of the above equation are bounded uniformly 
in $\theta$ and $n$. 
Thus, the ``local Doppler factors'' 
\Rf{eq:locDopfac} will be bounded, 
which implies the boundedness of the energy 
\Rf{eq:energy_time}.

\proofend

There is an interesting connection between 
the invariant densities of the system and 
the growth of the electromagnetic energy density.

Recall that if a density $\mu$ is invariant, 
$\mu(G(\theta)) = \mu(\theta) / G'(\theta)$. 
Hence, if the density $\mu$ never vanishes, 
$G'(\theta) = \mu(\theta) / \mu(G(\theta))$ 
and, therefore, 
$(G^i)'(\theta) = \mu(\theta) / \mu(G^i(\theta))$. 
Let us assume that there is only one characteristic 
passing through the space-time point $(t,x)$, 
and this characteristic is going to the right. 
Then, using the notations 
of Sec.~\ref{sec:sol_problem}, we can write the 
energy density at $(t,x)$ as [cf.~\Rf{eq:locDopfac}] 
\begin{equation}
\label{eq:inv_density} 
{\cal T}^{00} (t,x) = \displaystyle{
	\left[ 
	\frac{1-\differ{a}(\theta^{-}_{-n_{-}})}
			{1+\differ{a}(\theta^{-}_0)} \,
	\frac{\mu(G^{n_{-}}(\theta^{-}_{-n_-}))}
		{\mu(\theta^{-}_{-n_-})}
	\right]^2 
	{\cal T}^{00}(t_0,x_0^-)}	\,.
\end{equation}

In the general case 
[with two characteristics through $(x,t)$], 
one can use \Rf{eq:vect_pot} and \Rf{eq:energy} to 
prove the following result:

\begin{lem}
\label{cor:measure}
If a system has an invariant density $\mu$ which is bounded 
away from zero, then the electromagnetic energy density 
of a $C^1$ initial data is smaller than $C\mu^2$ for all times.
\end{lem}

In the cases that Herman's theorem applies, there is 
an invariant density bounded away from zero (and also bounded).
Hence, we conclude that there are values 
of the amplitude of mirror's oscillations for which 
the energy density of the field remains bounded. 
This set is typically a Cantor set interspersed 
with values for which the energy increases exponentially.

Some other results about the behavior of the energy 
with respect to time and parameters 
are obtained in~\cite{Dittrichetal1997}.

We call attention to the fact that \cite{Arnold1961} 
contains examples of analytic maps whose rotation numbers 
are very closely approximated by rationals 
and that are arbitrarily close to a rotation such that they 
preserve no invariant density and, therefore, are not smoothly 
conjugate to a rotation.

It is also known that for all rotation numbers one can construct 
$C^{2-\varepsilon}$ maps arbitrarily close to rotations with this 
rotation number and such that they do not preserve any invariant 
measures~\cite{HawkinsSchmidt1982}. 
It is a testament to the ubiquity of these maps that 
these questions were motivated and found applications 
in the theory of classification of $C^*$~algebras.


\subsection{The behavior for small amplitude and universality}
\label{sec:small_amplitude}

We note that, even if all the motions of the mirror 
lead to a circle map as in (\ref{eq:Fper}), it does 
not seem clear to us that all the maps of the 
circle can appear as 
$F$, $G$ for a certain~$a$.
This makes it impossible to 
conclude that the theory of generic 
circle maps applies directly to 
obtain conclusions  for a generic motion of 
the mirror. Of course, all the conclusions of 
the general theory that apply to all maps of 
the circle apply to our case. Those 
conclusions that require non-degeneracy assumptions
will need that we verify the assumptions.
Nevertheless, the very developed mathematical 
theory of generic or universal circle maps 
cannot be applied without caution 
to maps that appear as the result of generic 
or universal oscillations of the mirror.

One aspect that we have found makes a big difference 
with the generic theory is the situation 
where the mirror oscillates with small amplitude, i.e., 
$a_\varepsilon(t) = \bar a + \varepsilon \addition(t)$
with $\addition$ a periodic function 
of zero average and period $1$, and $\varepsilon\ll 1$. 
The first parameter, $\bar{a}$, is the average length 
of the resonator, while $\varepsilon=0$ is called 
the ``nonlinearity parameter'' for obvious reasons. 
If we denote by $F_{\bar{a},\varepsilon}$ 
and $G_{\bar{a},\varepsilon}$ 
the corresponding 2-parametric families of maps of  
the circle constructed according to (\ref{eq:Fexp}), 
then we have, for  three times differentiable families, 
\begin{eqnarray}
\label{eq:small_amp}
\FL
F_{\bar{a},\varepsilon}(t) &=& t + 2 \bar a 
+  2 \varepsilon \addition (t + \bar a)
+ 2 \varepsilon^2 \differ{\addition} 
	(t + \bar a) \addition (t + \bar a) 
+ O(\varepsilon^3)		\,,	\nonumber \\ [-2mm]
\FL						  \\ [-2mm]
\FL
G_{\bar{a},\varepsilon}(t) &=& t + 2 \bar a 
+  \varepsilon [ \addition (t) 
+   \addition  (t + 2  \bar a)  ]
+ \varepsilon^2 \differ{\addition} (t + 2 \bar a) 
	[ \addition  (t) + \addition  (t + 2 \bar a) ] 
+ O(\varepsilon^3)		\,.	\nonumber
\end{eqnarray}
Note that the term of order $\varepsilon$ has 
vanishing average.
As we will immediately show, 
this property causes that some well known 
generic properties of families of 
circle mappings do not hold for families 
of maps constructed as in \Rf{eq:Fexp}.

Indeed, if we consider the expressions 
for small amplitude developed in 
\Rf{eq:small_amp}, we can write the maps as 
$$
H_\varepsilon (t) = t + 2 \bar{a} + \varepsilon H_1(t) 
	+ \varepsilon^2 H_2(t) + O(\varepsilon^3)	\,.
$$
Since the conclusions of the theory of circle maps 
are independent of the coordinate system chosen, 
it is natural to try to choose a coordinate system 
where these expressions are as simple as possible. 
Hence, we choose $\heps(t):=t+\varepsilon\htil(t)$, 
a perturbation of the identity, and consider 
$h_\varepsilon^{-1}\circ\Heps\circ\heps$, 
which is just $\Heps$ in another system of coordinates, 
related to the original one by $\heps$. 
Then, up to terms of order $\varepsilon^3$, we have 
\begin{eqnarray}
\label{eq:transformed}
\FL
h_\varepsilon^{-1}\circ\Heps\circ\heps(t) 
&=& t  + 2 \bar{a} 
+ \varepsilon \left[ \htil(t) - 
		\htil(t+2\bar{a}) + H_1(t) \right]
		\nonumber\\[-2mm]
\FL		\\[-2mm]
\FL
&&\hspace{-15mm}+\varepsilon^2 \left\{ 
	\htil'(t+2\bar{a})\htil(t+2\bar{a}) 
	- \htil'(t+2\bar{a}) \left[\htil(t)+H_1(t)\right] 
	+ H'_1(t) \htil(t) + H_2(t) \right\} \,. \nonumber
\end{eqnarray}
We would like to choose $\htil$ in such a way that 
the $\varepsilon$ term is not present. 
Note that since 
$\int \htil(t+2\bar{a}) \,\der t = \int \htil(t) \,\der t$, 
this is impossible unless $\int H_1(t) \,\der t = 0$. 
When $\int H_1(t) \,\der t = 0$, $H_1$ is smooth and $2\bar{a}$ 
is Diophantine, a well-known result 
(see, e.g., \cite[Sec.\ XIII.4]{Herman1979}) 
shows that in such a case we can obtain 
one $\htil$ satisfying 
\begin{equation}
\label{eq:linearized}
\htil(t) - \htil(t+2\bar{a}) + H_1(t) = 0
\end{equation}
and $\overline{\htil}=0$. 
[Such $\htil$ is conventionally 
obtained by using Fourier coefficients. 
Note that in Fourier coefficients, \Rf{eq:linearized} 
amounts to 
$\widehat{\htil}_k (e^{2\pi i k 2 
	\bar{a}}-1) = \widehat{(H_1)}_k$. 
If $H_1$ is smooth, the Fourier coefficients decrease fast and 
if $2\bar{a}$ is Diophantine, then 
$(e^{2\pi i k 2 \bar{a}}-1)^{-1}$ does not grow too fast. 
For more details we refer to the reference above.]

Since for the functions $F_{\bar{a},\varepsilon}$ and 
$G_{\bar{a},\varepsilon}$ the term of order $\varepsilon$ 
has a zero average, we can transform these functions 
into lifts of rotations plus $O(\varepsilon^2)$. 
This implies, in particular, that their rotation number is 
$\tau(F_{\bar{a},\varepsilon}) = 
\tau(G_{\bar{a},\varepsilon}) = 2 \bar{a} + O(\varepsilon^2)$. 
One could wonder if it would be possible to continue the 
process and eliminate also to order~$\varepsilon^2$.

If we look at the $\varepsilon^2$ terms in \Rf{eq:transformed}, 
we see that $\overline{\htil'(t) \htil(t)} = 0$, 
and, when $\htil$ is chosen as in \Rf{eq:linearized}, 
$$
\htil'(t+2\bar{a}) [ \htil(t) + H_1(t) ] 
	= \htil'(t+2\bar{a}) \htil(t+2\bar{a})	\,,
$$
which also has average zero. 
Therefore, a necessary condition for the $\varepsilon^2$ term 
in $h_\varepsilon^{-1}\circ\Heps\circ\heps(t)$ to be zero is 
$\overline{H'_1(t)\htil(t)} + \overline{H_2(t)} = 0$. 

For the $F_{\bar{a},\varepsilon}$ in \Rf{eq:small_amp} 
we see that $F_2$ has zero average. 
Nevertheless, the term $F_1'(t) \htil(t)$ 
does not in general have average zero 
as can be seen in examples. Hence, we see that 
the rotation number indeed changes by an order which is 
$O(\varepsilon^2)$ and not higher in general. 
This property is not generic for families of circle maps 
starting with a rotation $2\bar{a}$ and it puts them 
outside of the universality classes considered in 
\cite{Shenker1982,Lanford1986}, etc., 
since the correspondence between rotation numbers and parameters 
is not the same.

According to the geometric picture of renormalization 
developed in \cite{Lanford1986}, 
the space of circle maps is divided into slices 
of rational rotation numbers, which are 
-- in appropriate sense -- parallel. 
In that language -- 
in which we think of families of circle maps 
as curves in the space of mappings -- 
the families of advance maps $F_{\bar{a},\varepsilon}$ 
and $G_{\bar{a},\varepsilon}$ (for fixed $\bar{a}$) 
have second order tangency to the foliation 
of rational rotation numbers rather than being 
transversal. 
Hence, the scaling predicted by universality theory 
should be true for $\varepsilon^2$ in place of $\varepsilon$. 
We have not verified this prediction, but we expect 
to come back to it soon.


\subsection{Schwarzian derivative 
	in the problem of moving mirrors}
\label{sec:schwarz}

Fulling and Davies 
\cite{FullingDavies1976} calculated 
the energy-momentum tensor in the two-dimensional 
quantum field theory of a massless scalar field 
influenced by the motion of 
a perfectly reflecting mirror 
(see also \cite{MostepanenkoTrunov1997}). 
They obtained that 
the ``renormalized'' vacuum expectation 
value of the energy density radiated 
by a moving mirror into initially empty space is
$$
{\cal T}^{00}(u) = - \frac1{24\pi} 
	\left[ \frac{F'''(u)}{F'(u)} - \frac32 
	\left(\frac{F''(u)}{F'(u)}\right)^2 \right]  \,,
$$
where $u=t-x$, and $F$ is related 
to the law of the motion of the mirror, 
$x=a(t)$, by~\Rf{eq:Fexp}. 
The right-hand side of this equation is 
nothing but (up to a constant factor) 
the Schwarzian derivative of $F$ 
-- a differential operator 
which naturally appears in complex analysis, 
e.g., it is invariant under 
a fractional linear transformation; 
vanishing Schwarzian derivative 
of a function is the necessary and 
sufficient condition that 
the function is fractional linear 
transformation, etc.
More interestingly, the Schwarzian derivative 
has been used as an important tool 
in the proof of several important theorems 
in the theory of circle maps 
-- see, e.g.,~\cite{Yoccoz1984,Herman1985}. 
In the light of the connection 
between the solutions of the wave 
equation in a periodically 
pulsating domain and the theory of 
circle maps it is not impossible 
that this is not just a coincidence.


\section{Conclusion}

Using the method of characteristics 
for solving the wave equation, 
we reformulated the problem of 
studying the electromagnetic field 
in a resonator with a periodically 
oscillating wall 
into the language of circle maps. 
Then we used some results of the 
theory of circle maps in order 
to make predictions about the 
long time behavior of the field. 
We found that many results in the 
theory of circle maps 
have a directly observable physical meaning. 
Notably, for a typical family of mirror motions 
we expect that the electromagnetic energy grows 
exponentially fast in a dense set 
of intervals in the parameters. 
Nevertheless, it remains bounded for all times 
for a Cantor set of parameters that has positive measure.

There are several advantages of the approach presented here. 
First, it allows us to understand better the time evolution 
of the electromagnetic field in the resonator and 
the mechanism of the change in the field energy. 
Second, the predictions are based 
on the general theory of circle maps 
so they are valid for any periodic motion of the mirror; 
let us also emphasize that our method is non-perturbative. 
Last, but not least, for a given motion 
of the mirror, one can easily make 
certain predictions about the behavior of the field by 
simply calculating the rotation number 
of the corresponding circle map, 
and without solving any partial differential equations.


\acknowledgments

This research was partially supported by N.S.F. grants.



\pagebreak

\pagebreak

\pagebreak

\pagebreak

\pagebreak

\pagebreak

\end{document}